&latex209
\documentstyle[12pt,aps,prb]{revtex}

\input psfig.tex

\begin{document}
\draft
\title{Relative energetics and structural properties of zirconia
             using a self-consistent tight-binding model}
\author{Stefano Fabris, Anthony T. Paxton and Michael W. Finnis}
\address{Atomistic Simulation Group, Department of Pure and Applied
Physics, Queen's University, \\ Belfast BT7 1NN, United Kingdom}
\date{\today} 
\maketitle
\begin{abstract}

We describe an empirical, self-consistent, orthogonal tight-binding model
for zirconia, which allows for the polarizability of the anions at dipole
and quadrupole levels and for crystal field splitting of the cation $d$
orbitals.  This is achieved by mixing the orbitals of different symmetry
on a site with coupling coefficients driven by the Coulomb potentials up
to octapole level. The additional forces on atoms due to the
self-consistency and polarizabilities are exactly obtained by
straightforward electrostatics, by analogy with the Hellmann-Feynman
theorem as applied in first-principles calculations. The model correctly
orders the zero temperature energies of all zirconia polymorphs. The Zr-O
matrix elements of the Hamiltonian, which measure covalency, make a
greater contribution than the polarizability to the energy differences
between phases.  Results for elastic constants of the cubic and
tetragonal phases and phonon frequencies of the cubic phase are also
presented and compared with some experimental data and first-principles
calculations. We suggest that the model will be useful for studying
finite temperature effects by means of molecular dynamics.

\end{abstract}
\pacs{31.15.Ar,71.15.Fv,81.30-t}

\section{Introduction}

Solid solutions of zirconia (ZrO$_2$) containing other oxides are among
the major representatives of modern ceramic materials. The wide range of
applications, including traditional structural refractories, fuel cells
and electronic devices such as oxygen sensors,~\cite{Heuer81,Claussen84}
testifies to the technological importance of zirconias. Different
divalent and trivalent oxides are added to ZrO$_2$ in order to improve
its thermomechanical properties, and charge-compensating vacancies are
thereby introduced on the anion sublattice. The macroscopic effects
associated with the impurities are very well
known,~\cite{Grain67,Scott75,Stubican81} but a microscopic model which
gives a theoretical interpretation is still missing. As a preliminary
step, this paper provides a physical picture of the crystal
thermodynamics of pure zirconia, combining the results of first
principles density functional and semiempirical Tight Binding (TB)
calculations.

\begin{figure}
\caption{Cubic and tetragonal structures of ZrO$_2$. Light and dark
circles denote oxygen and zirconium atoms respectively. Arrows represent
the structural instability of the oxygen sublattice along the $X_2^-$
mode of vibration.}
\label{cellct}
\centerline{\psfig{file=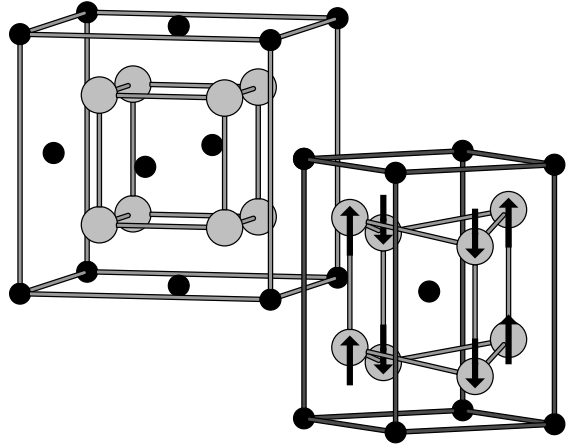,height=6cm,angle=-90} }
\end{figure}

Zirconia has three zero-pressure polymorphs; these have cubic
($c$), tetragonal ($t$) and monoclinic ($m$) symmetry. The high
temperature $c$ phase \cite{Aldebert85,Ackermann77} ({\it Fm}3{\it m}) is
stable between 2570 K and the melting temperature of 2980 K.  The $t$
structure \cite{Howard88,Teufer62} ($P4_2/nmc$), which is stable between
1400 and 2570 K, is closely related to the $c$ one: the internal degree
of freedom $\delta$ shifts the oxygen ions away from the centrosymmetric
positions along the $X_2^-$ mode of vibration (Figure~\ref{cellct}) and
forces the $c/a$ ratio of the unit cell to adjust. Below 1400 K the
low-symmetry $m$ phase \cite{Adam59,McCullough59,Smith65} ($P2_1/c$) is
thermodynamically stable.

Besides its technological implications, the relationship between these
structures is of fundamental interest. The mechanisms of the phase
transformations, the effects of impurities and vacancies on
them, and their relationship to the nature of the bonding still require
explanation, and this may shed light on the properties of other, more
complex oxides.

The crystal structure of purely ionic bonded materials can be determined
on the basis of radius-ratio rules,~\cite{Kingery60} based purely on
electrostatic arguments. Because of the small size of the Zr$^{4+}$
ions, these rules place ZrO$_2$ on the border between the 8-fold
coordinated fluorite structure and the 6-fold coordinated rutile one
($P4_2/mnm$). The radius-ratio is too blunt a tool to account for the
absolute stability of the unique 7-fold coordinated $m$ structure.


The classical empirical models of zirconia are based on the {\it a
priori} assumption of its {\it ionicity}. Empirical approaches like the
Shell Model (SM) or the Rigid Ion Model (RIM) described the
structural,~\cite{Dwivedi90} dynamical \cite{Mirgorodsky97,Cormack90}
and transport \cite{Li95,Shimojo92-I,Shimojo92-II} properties of the
phases on which they were parameterized, but failed to predict the
absolute stability of the $m$ structure.  The most detailed of such
models was developed by Wilson {\it et al.},~\cite{Wilson96zirc} whose
environment-dependent Compressible and Polarizable Ion Model (CIM-DQ)
demonstrated the importance of the anion polarizabilities at both dipole
and quadrupole levels on the energetics of zirconia. However, further
calculations~\cite{Laurea97} carried out with this model revealed that
even though it predicted the correct energy ordering of the $c$, $t$ and
$m$ phases, it predicted that the rutile structure should be even more
stable, and this phase is never observed experimentally in zirconia.

The experience gained with the CIM-DQ model suggests that a successful
empirical model of zirconia should describe the effects of the atomic
polarization, but should also go beyond a purely ionic description of the
bonding. The partial covalent character of zirconia has already been
postulated \cite{Ho82} and is evident from electronic structure
calculations based on density functional theory. In this paper we further
investigate the recently proposed polarizable self-consistent tight
binding (SC-TB) model \cite{Finnis97,Finnis98,Schelling98} which combines
the physical concepts of covalency, ionicity and polarizability. Using
the SC-TB model we are drawn to the conclusion that the covalent
character of the Zr-O bond makes a significant contribution to the
relative energetics of different structures, which would explain the
limited predictive power of the previous ionic models.

There have been several previous approaches to analyzing the structural
and electronic properties of zirconia.  Boyer and Klein \cite{Boyer85b}
used the APW method to derive pair potentials with which to investigate
the equation of state of the $c$ phase. Cohen {\it et al.} \cite{Cohen88}
calculated the relative energetics and the elasticity using the Potential
Induced Breathing (PIB) method based on the Gordon-Kim
approach. Zandiehnadem {\it et al.}  \cite{Zandiehnadem88} studied the
electronic structure with a first principles LCAO method. The FLAPW
calculations of Jansen \cite{Jansen91} predicted for the first time the
correct energetic ordering between the $c$ and $t$ structures at zero
absolute temperature, identifying the double well in the potential
energy that governs their relative stability. The double well was subsequently
confirmed by {\it ab initio} Hartree-Fock (HF)
calculations,~\cite{Orlando92,Stefanovich94} but these did not predict
the stability of the $m$ structure over the $t$ one. Only the very recent
Density Functional Theory (DFT) calculations
\cite{Kralik98,Stapper99,Jomard99} consistently reproduce the relative
energetics of the three zirconia polymorphs at 0 K.

The plan of the present paper is as follow. In Section \ref{TB model} we
describe the model used in the calculations, the inclusion of the atomic
polarizability in the TB framework and the parameterization procedure. A
preliminary account of this work has been published.~\cite{Finnis98} We
have made DFT calculations of band structures of the simple structures
for this purpose, using a new full-potential, linear muffin tin orbital
method (NFP-LMTO). The predictive power of the new model is tested
against the DFT calculations in Section \ref{en-vol}, where we study the
relative energetics of zirconia. Section \ref{c-t dist} focuses on the
relationship between the $c$ and $t$ structures: the Landau theory of
phase transformation is used to interpret the results of the static
calculations. In Section \ref{distortions}, we explore the elastic and
the vibrational properties of the high symmetry phases. The results are
summarized in the concluding Section.

\section{The Tight Binding Model}
\label{TB model}

\subsection{Including polarizabilities in TB}

In the TB approximation the crystal wave function can be expressed as a
linear combination of atom-centered orbitals which we denote $\left. \mid
{\bf R} L \right>$: 

\begin{equation}
\left. \mid \Psi^{n {\bf k}} \right> = \sum c_{{\bf
R} L}^{n {\bf k}} \left. \mid {\bf R} L \right>.
\end{equation}
$L$ is a composite angular momentum index $L=\left(\ell,m \right)$ of the
atomic orbital centered on the site whose position is {\bf R}, $n$ and
${\bf k}$ are the band and ${\bf k}-$vector indices of the single
particle wave function. For the purpose of derivation, we express the
local orbitals as a product of a radial function and a real spherical
harmonic
\begin{equation}
\left< {\bf r} \! \mid\! {\bf R}  L \right> = f_{{\bf R}\ell}(\mid \!
{\bf r} - {\bf R} \! \mid ) \, Y_{L}({\bf r}- {\bf R}), \label{loc-orb}
\end{equation} 
although in our {\it empirical} TB scheme the explicit functional forms
of the radial wave functions are not required. To simplify the notation,
we will frequently suppress the site index ${\bf R}$, in which case one
can take it we are referring to an atom at the origin and ${\bf r}$ is a
small vector in its neighborhood.


The total Hamiltonian ${\cal H}$ can be expressed as a sum of two terms,
${\cal H} = {\cal H}^0 + {\cal H}^\prime$. In traditional Self-Consistent
(SC) TB, ${\cal H}^0$ contains both on-site and inter-site terms. The
on-site terms are diagonal in $L$, and are often taken as Hartree-Fock
term values of the isolated atoms. The inter-site terms are the bonding
integrals. The additional part of the Hamiltonian, ${\cal H}^\prime$, is
diagonal in {\bf R} and $L$ in the traditional approach (Majewski and
Vogl \cite{Majewski86,Majewski87}). It controls the charge redistribution
between neighboring sites which results from the balance between the
opposite effects due to the on-site Coulomb repulsion (Hubbard $U$) and
Madelung potentials.


What is missing in the previous model is the effect of the crystal
fields on the valence electrons, {\it i.e.} the atomic {\it
polarizability}. In a preliminary account of this work \cite{Finnis98}
we indicated how to include the polarization effects in a
SC-TB formalism by adding off-diagonal terms ${\cal H}^\prime_{{\bf
R}L{\bf R}L^\prime}$ to the on-site blocks of the Hamiltonian. Here we
describe how we make that extension.

If we assume the on-site charge distribution to be localized, then its
total multipole moment $Q_L$ has a monopole contribution from the ionic
core charge and a multipole (including monopole) contribution from the
valence charge:

\begin{equation}
Q_{L} = Q^i \, \delta_{L0} + Q^e_{L}. 
\label{tmpol}
\end{equation}

As Stone \cite{Stone96} points out, the electronic multipole moment on a
site is the expectation value of the operator 
\begin{equation}
 \hat{Q}_L^e=e \, \hat{r}^\ell Y_L({\bf \hat{r}}), \label{stone}
\end{equation}
where $e$ is the charge of the electron. Neglecting inter-site terms like
$\left<{\bf R^\prime} L^\prime \left| \hat{Q}^e_{{\bf R} L} \right| {\bf
R^{\prime\prime}} L^{\prime\prime} \right>$ for ${\bf R^\prime},{\bf
R^{\prime\prime}} \ne {\bf R}$, the definition of the on-site multipole
moment is therefore:

\begin{equation}
Q_L^e \equiv \sum_{L^\prime L^{\prime\prime}} \sum_{n {\bf k}}^{\rm occ.}
c_{L^\prime}^{n {\bf k}} c_{L^{\prime\prime}}^{n {\bf k}} \left<L^\prime
\left| \hat{Q}^e_L \right| L^{\prime\prime} \right>. \label{mpol}
\end{equation}
By invoking equations (\ref{loc-orb}) and (\ref{stone}), the last factor
of Eq.(\ref{mpol}) can be expressed as a product of two quantities, the
Gaunt coefficients $C_{L^\prime L^{\prime\prime}L}$, which dictate the
selection rules, and the integrals $\Delta_{\ell^\prime
\ell^{\prime\prime} \ell}$, which will be new parameters of the model:


\begin{eqnarray}
\left< L^\prime \left| \hat{Q}^e_L \right| L^{\prime\prime} \right> & = &
e \,\, {\Delta_{\ell^\prime \ell^{\prime\prime} \ell}} \,\, {C_{L^\prime
L^{\prime\prime} L}} \label{product} \\
{C_{L^\prime L^{\prime\prime} L}} & = &  \int \!\! Y_{L^{\prime}}
Y_{L^{\prime \prime}} Y_L \, \, d{\Omega} \label{gaunt} \\  
{\Delta_{\ell^{\prime} \ell^{\prime\prime} \ell}} & = & \int \!\!
f_{\ell^{\prime}}(r) f_{\ell^{\prime\prime}}(r) r^{\ell+2} dr \, , \label{delta} 
\end{eqnarray} 
where $d \Omega$ stands for the element of solid angle $\sin \theta \,
d\theta \, d\phi$. The r$\hat{\rm o}$le of the Gaunt coefficients, which
depend on the angular part of the wave function only, is to select the
term with symmetry $L$ arising from the coupling of the on-site orbitals
$L^\prime$ and $L^{\prime\prime}$. The $\Delta$ parameters, depending on
the radial part of the wave function, determine the magnitude of the
coupling. The substitution of Eq.(\ref{product}) in Eq.(\ref{mpol})
defines the multipole moment of symmetry $L$ on the site {\bf R}.

Having defined the on-site multipole moments, we can calculate the fields
which they generate on all the lattice sites. The derivation uses
standard results from classical electrostatics. The electrostatic
potential is expanded in partial waves about the site:
\begin{equation}
V({\bf r}) = \sum_{L} V_L \, r^\ell Y_L({\bf r}),
\end{equation}
where, using the Poisson equation,
\begin{equation}
V_{L} = 4\pi  \sum_{{\bf R^\prime} \neq {\bf 0}} \sum_{L^\prime}
\tilde{B}_{L L^\prime} \left({\bf R^\prime} \right) 
Q_{{\bf R^\prime}L^\prime}, \label{pot}
\end{equation}
and
\begin{eqnarray}
\tilde{B}_{L L^\prime} \left({\bf  R} \right)  
& = & \frac{4\pi }{(2\ell+1)!! \, (2\ell^\prime
+1)!!} \\ 
& & \times  \sum_{L^{\prime \prime}} 
 \,
\frac{ (-1)^{\ell^\prime} \, (2\ell^{\prime \prime}-1)!!}{
\left| {\bf  R} \right|^{\ell^{\prime \prime} +1}} 
\, Y_{L^{\prime \prime}}({\bf  R}) \, C_{L^{\prime \prime} L^\prime L }  . \nonumber
\end{eqnarray}
The sum over $L^{\prime\prime}$ is restricted to the values for which
$\ell^{\prime\prime} = \ell + \ell^\prime$; $\tilde{B}_{L L^\prime}$ are
proportional to the well known LMTO-ASA structure
constants.~\cite{Andersen84} The component of electrostatic potential
$V_L$ couples different orbitals on a site giving the matrix elements:
\begin{equation}
\left< L^\prime \! \mid \! {\cal H}^\prime \! \mid \! L^{\prime\prime}
\right> = \sum_L \,   V_L \, \, \Delta_{\ell^{\prime} \ell^{\prime\prime}
\ell} \, \, C_{L^\prime L^{\prime\prime}L} .
\end{equation}

The diagonal elements of the Hamiltonian are adjusted by using a single
Hubbard $U$ in the standard way, which adds a term $U \,\delta N_{{\bf
R}\ell}$ to each diagonal matrix element. The quantities $\delta N_{{\bf
R}\ell}$ are the changes in the electronic charge projected onto a site and
orbital compared to the input, non-self-consistent charge. We use the
standard Mulliken projection. Finally the Schr\"odinger equation is
solved using a self-consistent iterative procedure with charge mixing to
obtain the coefficients $c_{{\bf R} L}^{n {\bf k}}$ and hence the
multipoles.

It is useful to step back at this point and compare the above model with
the Hohenberg-Kohn-Sham (HKS) one, whose exchange and correlation energy
functional $U^{xc}[n]$ has been expanded to second order in the electron
density n({\bf r}):~\cite{Foulkes89}

\begin{eqnarray}
   U^{\rm HKS} & = &  \sum_{n, {\bf k}} ^{\rm occ} \left< \Psi^{n
{\bf k}} \mid {\cal T}_S + V_0^{xc} + V_0^{\rm H} + V_0^i \mid \Psi^{n {\bf k}} \right>
\label{HKS} \\
& & + U^{xc}[n_0] - \int \!\! V^{xc}_0  \, n_0 \, d {\bf r}
 - U^{\rm H}[n_0] + U^{ii} \nonumber \\
& & + \frac{1}{2} \int \!\!\!\! \int \!\!\left(
\frac{e^2}{\mid \! {\bf r-r^{\rm \prime}} \!\! \mid} +
\left. \frac{\delta^2 U^{xc}}{\delta n \, \delta n^\prime}
\right|_{n=n_0} \right) \, \delta n \, \delta n^\prime \, d{\bf r}\, d{\bf
r^{\rm \prime}}. \nonumber
\end{eqnarray}
$n_0$ denotes a reference electron density, which we will consider as a
superposition of spherical ionic charges; ${\cal T}_S$ is the kinetic
energy operator of the non-interacting electron gas, $V_0^{xc}$,
$V_0^{\rm H}$ and $V_0^i$ are the exchange and correlation, Hartree and
ionic potentials calculated at the reference charge $n_0$; $\delta n$
denotes the deviation from that reference ($\delta n = n - n_0$) and
${n^\prime}$ refers to the electron density at ${\bf r^\prime}$. $U^{\rm
H}$ and $U^{ii}$ are respectively the Hartree and the ion-ion
electrostatic energies.

Without the last term, this is simply the Harris-Foulkes functional.  It
generates a non-self-consistent TB model in which the first term is the
sum of the eigenvalues while the second is a sum of pair
potentials.~\cite{Sutton88} If the last term is included, the total energy
must be minimized iteratively, and the last term now provides the
self-consistency correction to the Kohn-Sham Hamiltonian.

The last line of Eq.(\ref{HKS}) represent the Hartree energy of the
deviation from the reference charge, $U^{\rm H}[\delta n]$, and the
second order term of the $U^{xc}$ Taylor expansion. We can identify this
term in our SC-TB model as follows:
\begin{eqnarray}
\frac{1}{2} \int \!\!\!\! \int \!\!\left(
\frac{e^2}{\mid \! {\bf r-r^{\rm \prime}} \!\! \mid} +
\left. \frac{\delta^2 U^{xc}}{\delta n \, \delta n^\prime}
\right|_{n=n_0} \right) \, \delta n \, \delta n^\prime \, d{\bf r}\, d{\bf
r^{\rm \prime}}
\equiv \\
\equiv  \frac{1}{2}  \sum_{{\bf R}L} \left( U \, \delta N^2_{{\bf R}\ell} + 
 Q_{{\bf R}L}\, V_{{\bf R}L} \right) \, . \nonumber
\end{eqnarray}

Our total energy in the SC-TB model is therefore
\begin{eqnarray}
U^{\rm TB} & = & \sum_{n, {\bf k}}^{\rm occ} \left< \Psi^{n {\bf k}} \mid
{\cal H}^0 \mid \Psi^{n {\bf k}} \right> + U^{\rm pair} \nonumber \\ & &
+ \frac{1}{2} \sum_{{\bf R}L} \left( U \, \delta N^2_{{\bf R}\ell} + Q_{{\bf
R}L} \, V_{{\bf R}L} \right) \label{UTB}
\end{eqnarray}

It can be verified that, by minimizing the above expression
with respect to the expansion coefficients in the wave functions,
we recover the Schr\"odinger equation with the SC-TB Hamiltonian. 

Calculation of the forces on the ions is very straightforward once we
have the self-consistent wave functions and multipoles. For if an ion is
moved a small distance $\delta {\bf R}$, there is no change in total
electronic energy to first order in the $\delta c_{{\bf R} L}^{n {\bf
k}}$.  Therefore we can calculate the force due to the change in the
first term of (\ref{UTB}) by the conventional formulae, using the
derivatives of the non-self-consistent Hamiltonian matrix elements (see
following section).  In calculating the forces due to the last term of
(\ref{UTB}) we can hold the multipoles fixed and use standard
electrostatics. There is no contribution to the forces from the on-site
energy containing $U$. The simple form of these results for the forces in
TB is a direct analogy with the application of the Hellmann-Feynman
theorem in DFT.

\subsection{Parameterization}

Each parameter of the model has been adjusted to the results of NFP-LMTO
calculations, details of which are specified in the previous work on
zirconia.~\cite{Finnis98} Our TB description of zirconia uses a minimal
basis of atomic orbitals. The oxygen atoms are modelled with $2p$ and
$3s$ orbitals and with a fixed core charge of +4, while on the zirconium
atoms there are $4d$ orbitals and a core charge of +4. The purpose of the
$3s$ orbital on the oxygen is twofold: to allow an extra degree of
freedom for polarization, which is otherwise restricted to charge
transfer between its $2p$ orbitals, and to better reproduce the structure
of the conduction bands.

A repulsive Born-Mayer pair potential $U^{\rm pair}$ has been chosen
in order to reproduce the lattice parameter and the bulk modulus of the
$c$ phase. Only the first Zr-O coordination shell has been included in
this interaction.

\begin{figure*}
\caption{Band structure of cubic zirconia. In all the panels, starting
from the bottom it is possible to identify the oxygen 2$p$ valence bands
and the unoccupied zirconium 4$d$ bands which are partly hybridized with
the oxygen 3$s$ one. The large crystal field splitting of the 4$d$ bands
predicted by the LDA calculation (c) is reproduced with the SC-TB model,
(a) and (b), when the $\Delta_{ddg}$ parameter is included.}
\label{bndstr}
\centerline{\psfig{file=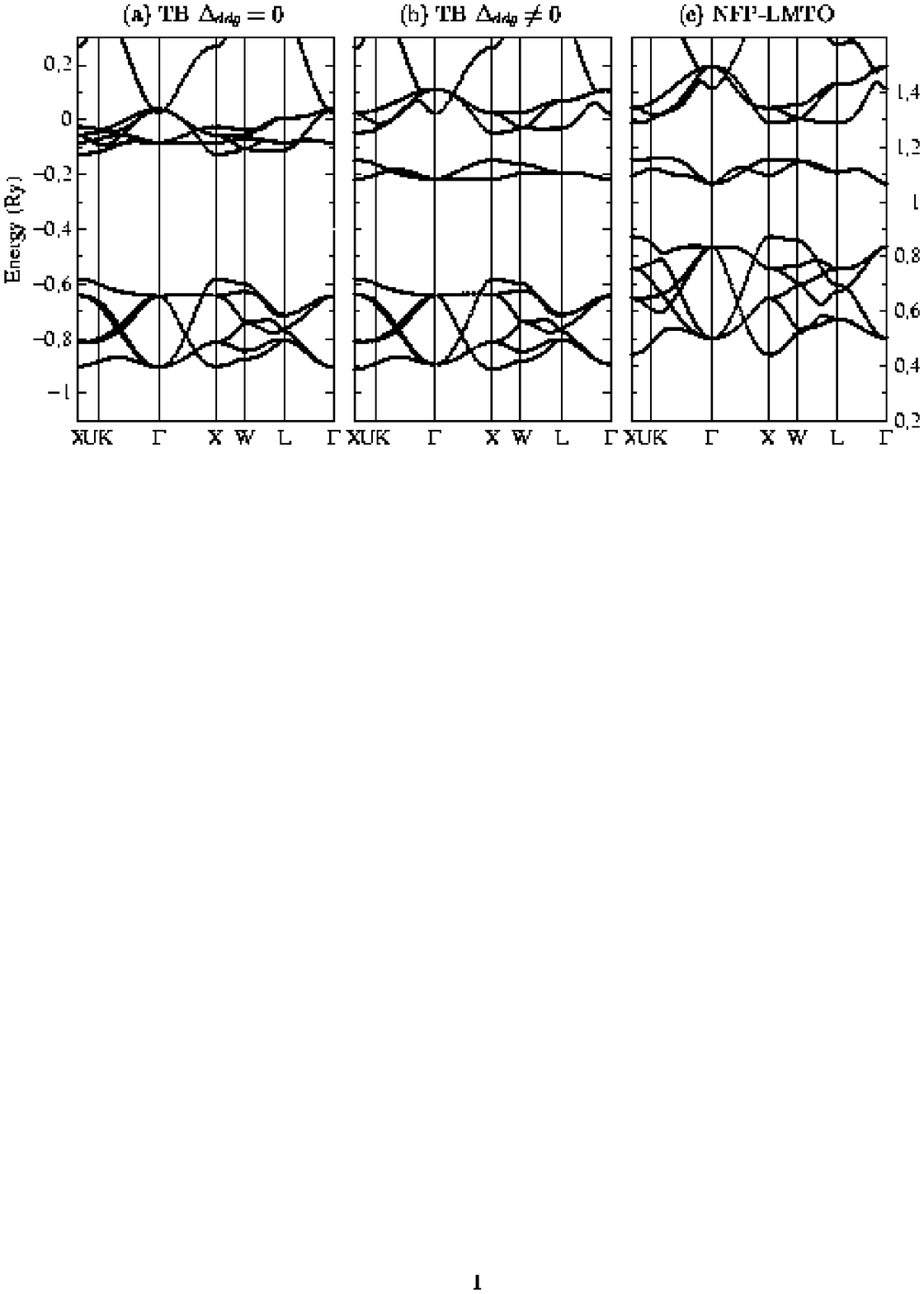,width=15cm,angle=0} }
\end{figure*}

The Hamiltonian ${\cal H}^0$ has been adjusted to the {\it ab initio}
electronic structure of the $c$ phase shown in Figure~\ref{bndstr} (c).
We chose the Goodwin-Skinner-Pettifor \cite{Goodwin89} distance
dependence of the 10 hopping integrals involved. The Hubbard $U$ have
been fixed to 1 Ry. The parameters of the SC-TB model are collected in
Table~\ref{param}.

The basis set chosen reduces the number of symmetry-allowed $\Delta$
parameters to 4: $\Delta_{spp}$, $\Delta_{ppd}$, $\Delta_{ddd}$ and
$\Delta_{ddg}$. The first two refer to the $s$ and $p$ orbitals of oxygen
ions, the last two to the $d$ orbitals on the zirconium.

In the highly symmetric $c$ structure the first non spherical terms of
the potential $V_L$ on the cation and anion sites have $g$ and $f$
symmetry respectively. The latter cannot interact with the oxygen
orbitals, the former splits the energetic levels of the zirconium $d$
orbitals and $\Delta_{ddg}$ determines the magnitude of the energy
splitting $\delta \epsilon$. Cubic crystal field theory
\cite{Stoneham75} predicts the proportionality between $\delta
\epsilon$ and the radial distribution of charge $< \! \! r^4 \! \! >$
which is the definition of $\Delta_{ddg}$ given in Eq.(\ref{delta}).
Figure~\ref{bndstr} (a) and (b) shows the effect of the $\Delta_{ddg}$
polarization term on the band structure of the $c$ phase: the
splitting of the $d$ bands could not be captured with the SC-TB without
the polarizability parameters. Reasonable values of the
$\Delta_{ddd}$ parameter have no significant effect on any physical
properties studied here, therefore we set it to zero.

Less symmetric structures are necessary to parameterize the remaining
$\Delta$'s. In the rutile phase, the $\ell=3$ component of the crystal
field acting on the oxygen ions splits the $p$ levels. Consequently, it
contributes to the width of the $2p$ band: this effect is controlled by
$\Delta_{ppd}$ which we adjust to match the {\it ab initio} band
structure of the rutile phase. The last term $\Delta_{spp}$ has been
chosen in order to reproduce the depth of the double well in the
potential energy of the $t$ structure.

\section{Energetics of Bulk Phases}

\subsection{Energy-Volume curves}
\label{en-vol}

\subsubsection{Zero-pressure phases}

\begin{figure}
\caption{SC-TB (top) and NFP-LMTO (bottom) Energy-Volume data for the cubic
(c), tetragonal (t), and monoclinic (m) phases of zirconia fitted with
Murnaghan equation of states.}
\label{envol}
\centerline{\psfig{file=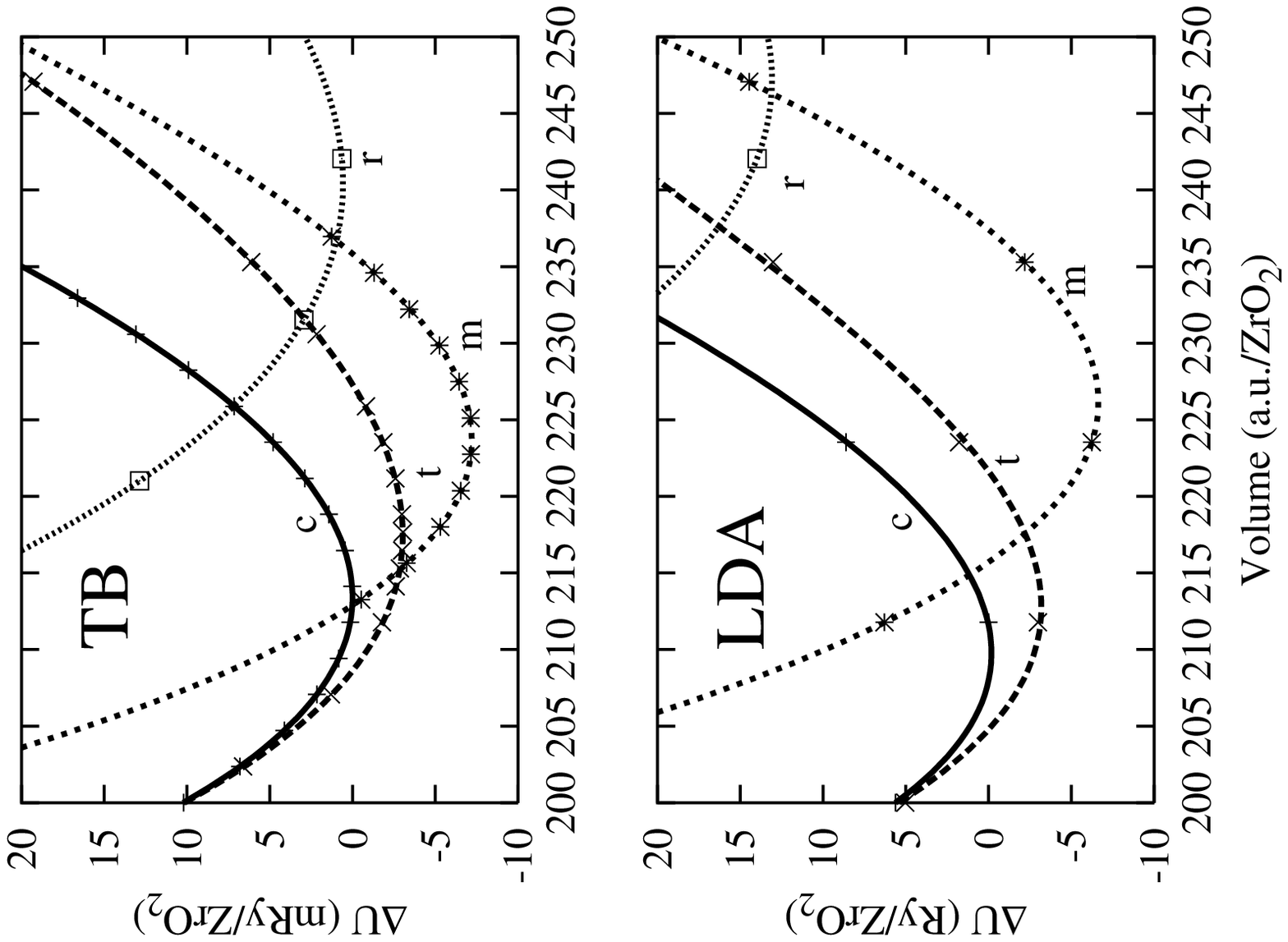,width=8.5cm,angle=-90}} 
\end{figure}

The predictive power of the polarizable TB model has been investigated
by comparing its results with NFP-LMTO calculations. The Energy-Volume
curves calculated with the two methods are shown in
Figure~\ref{envol}. Each energy value involved the full relaxation of
all the degrees of freedom of the structures.

The $c$ and the $t$ phases were used in the parameterization
procedure, therefore there is automatic agreement of the two methods
for these crystal structures. The true prediction of the model is the
absolute stability of the monoclinic phase. This indicates  the
transferability of the parameters between the phases. 

The rutile phase, which is not experimentally observed, has been included
in the study because further calculations with the CIM-DQ
\cite{Laurea97,MRS99} model predicted the rutile phase to be more stable
than the monoclinic one. Figure~\ref{envol} shows that the SC-TB model
does not suffer from this problem, although the relative energy of the
rutile phase is less than with the DFT. To our knowledge, the SC-TB is
the first semi-empirical model which reproduces the correct ordering of
these polymorphs at zero temperature, including the stability of the $m$
phase.

Table~\ref{strucpar} summarizes the structural properties calculated with
the NFP-LMTO method and with the polarizable SC-TB model, comparing them
with other theoretical and experimental works. The $c$ and $m$ lattice
parameters are referred to the 12-atoms unit cell, while the $t$ ones
are given in terms of the 6-atoms unit cell. A comparison of the energy
differences between the phases of zirconia calculated with different
methods is given in Table~\ref{energy}.

\subsubsection{High-Pressure phases}

Under pressure, the low temperature $m$ phase transforms to an
orthorhombic structure, known as ortho I ($o_I$), whose crystallography is
still controversial. X-ray diffraction analysis\cite{Kudoh86,Suyama85}
suggests it belongs to the $Pbcm$ space group while neutron diffraction
studies\cite{Ohtaka90,Howard91} propose the $Pbca$ space group. We
carried out the calculations using the latter structure. The phase
transition pressure strongly depends on the state of the sample and is
believed to be between 3 and 6 GPa.~\cite{Liu80,Block85,Ohtaka91} A
second pressure-induced phase transition is observed around 15
GPa,~\cite{Ohtaka91} where the $o_I$ transforms to the orthorhombic phase
termed ortho II ($o_{II}$). The latter is isostructural to cotunnite
(PbCl$_2$) and belongs to the $Pnam$ space group.~\cite{Ming85} The
pressure increases the coordination number of the zirconium atoms from 7
to 9. 

A comprehensive first-principles study of the two orthorhombic phases has apparently not
yet been made: Stapper {\it et al.}\cite{Stapper99} studied the $o_I$
structure only, while Jomard {\it et al.}\cite{Jomard99} focused on the
$o_{II}$ phase. 

The atomic environment of the high pressure phases is completely
different to that of the $c$ and $t$ phases used in the
parameterization of the TB model, therefore these orthorhombic structures
provides a severe benchmark for the transferability of the TB
parameters.

\begin{figure}
\caption{Energy-Volume curves for the monoclinic (m) and orthorhombic
($o_I$ and $o_{II}$) phases calculated with the TB model.}
\label{ortenvol}
\centerline{\psfig{file=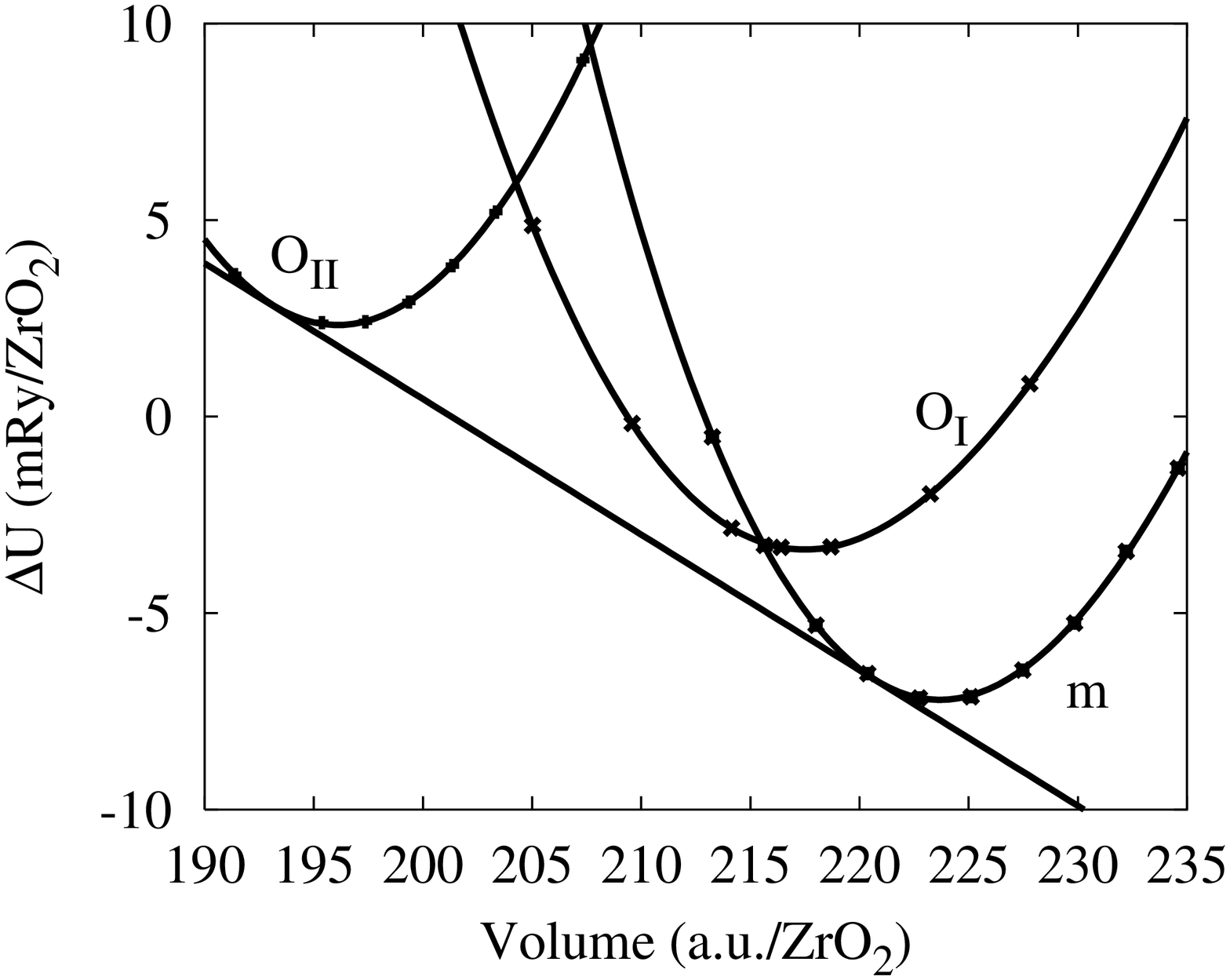,width=8.5cm,angle=-90}}
\end{figure}

The energy ordering of the phases predicted by the TB model is

\[ U^m < U^{o_I} < U^t < U^c < U^{o_{II}}, \] 
which is the same as we obtain by combining the results of
Refs. \onlinecite{Stapper99} and \onlinecite{Jomard99}. The numerical
values of the energy differences are summarized in Table~\ref{energy} and
compare reasonably well with the {\it ab initio} results. The
Energy-Volume curves of the orthorhombic phases are shown in
Figure~\ref{ortenvol}: all the degrees of freedom were fully relaxed and
their values are collected in Table~\ref{parort}.

Although the TB model predicts the correct relative energetics
of the phases, it is not capable of describing the subtle
pressure-induced phase transformation $m \leftrightarrow
o_I$. Figure~\ref{ortenvol} shows the common-tangent between the
$m$ and the $o_{II}$ phases. As the pressure is increased,
the model misses the correct sequence of the phases,
predicting a $m \leftrightarrow o_{II}$ pressure-induced phase
transformation at 5 GPa. 

\subsection{Cubic versus Tetragonal Phases}
\label{c-t dist}

\subsubsection{Static calculations}

The relationship between the cubic and the tetragonal phases is governed
by a volume dependent double well in the potential energy. Since the FLAPW
calculation of Jansen \cite{Jansen88,Jansen91} who predicted it first,
the double well has been confirmed by several other {\it ab initio} calculations
and it is now well established.

In this section we analyze the nature of the 0 K energy surface by
combining the information gained using two very different approaches: the
NFP-LMTO method and the polarizable TB model. The qualitative and
quantitative agreement between the results of the two calculations, shown
in the previous section, entitles us to use the physical picture provided
by the simpler model to interpret the {\it ab initio} results.

Starting from the $c$ phase, the $t$ structure can be obtained by
continuously stretching the unit cell along  the $c$  crystallographic
direction and by displacing the oxygen columns by $\delta$ along the
tetragonal axis according to the $X_2^-$ mode of vibration
(Figure~\ref{cellct}). We calculated the total energy of the crystal
using the two methods, for different values of ($\delta$, $c/a$) at
several volumes.

\begin{figure*}
\caption{SC-TB cohesive energy vs. tetragonal distortion $\delta$: volume and
$c/a$ dependence. (a) Single well at V$_1$=198 a.u/ZrO$_2$; (b) Double
well at V$_2$=218 a.u/ZrO$_2$.}
\label{well}
\centerline{\psfig{file=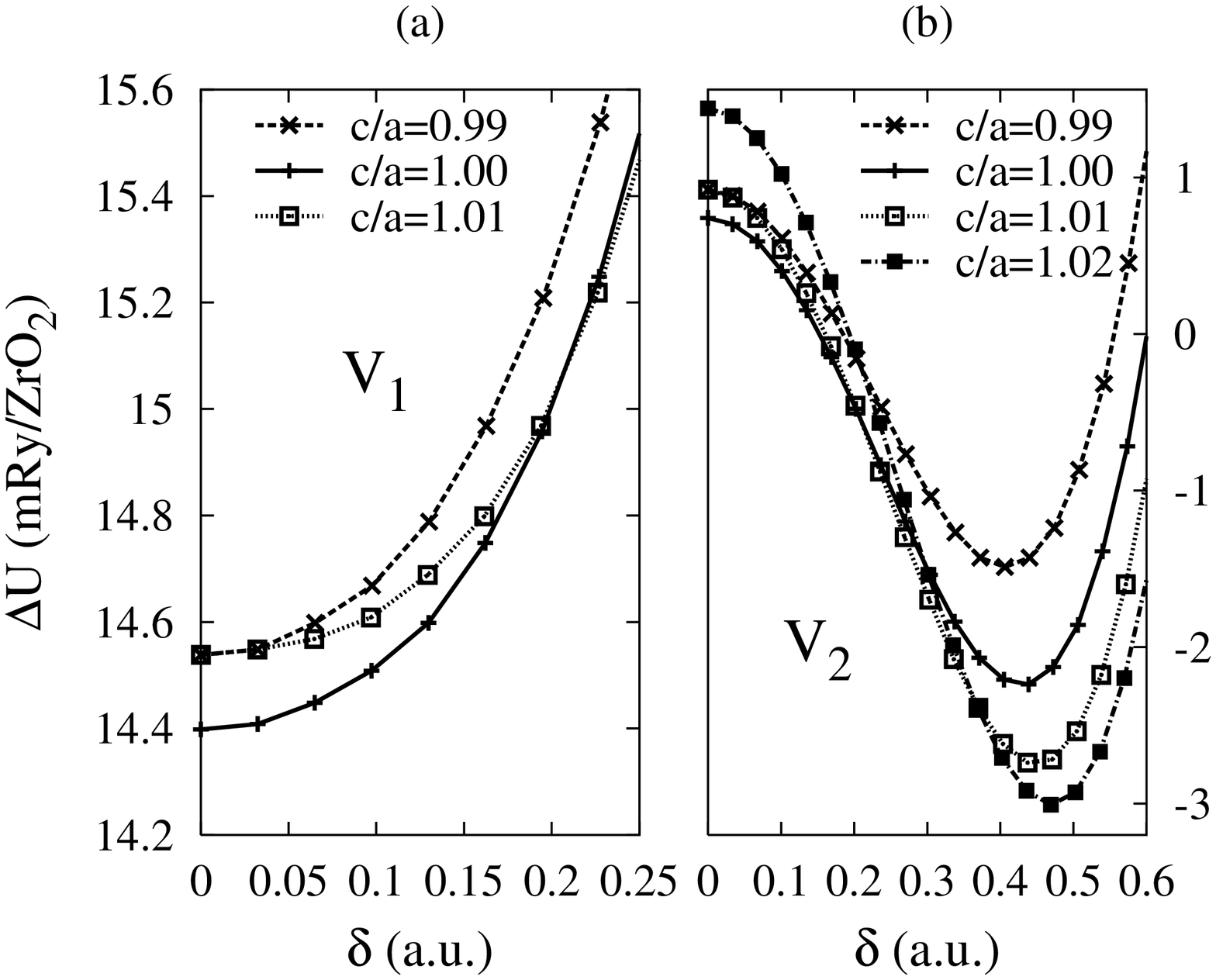,width=8.5cm,angle=-90} }
\end{figure*}

The energy curve exhibits a single well or a double well structure
depending on the specific volume. At small volumes, V$_1$, the tetragonal
distortion is energetically unfavored and the equilibrium structure is
cubic (Figure~\ref{well}). When the cubic phase is stable, there is no
distinct metastable tetragonal phase with which to compare its energy, so
the energies of the two phases merge. At larger volumes, V$_2$, a
structural instability appears and the $c$ structure spontaneously
distorts to the $t$ one (Figure~\ref{well}).

The curvature of the energy surfaces is related to the phase transition
mechanism. It is clear from Figure~\ref{well} that $\frac{\partial^2E}{\partial \eta^2}$ is positive, while $\frac{\partial^2 E}{\partial
\delta^2}$ is negative: this suggests that the phase transition is
driven by the $\delta$ instability and that the adjustment of the $c/a$
ratio is a secondary effect. The coupling between these two order
parameters will be further discussed when we interpret the double
well using Landau Theory.

\begin{figure}[]
\caption{Phonon dispersion curves of cubic zirconia in the [100]
direction. Closed circles are TB calculations, dashed lines are guides to
the eye. Note the imaginary frequency of the $X_2^-$ mode of vibration.}
\label{phon}
\centerline{\psfig{file=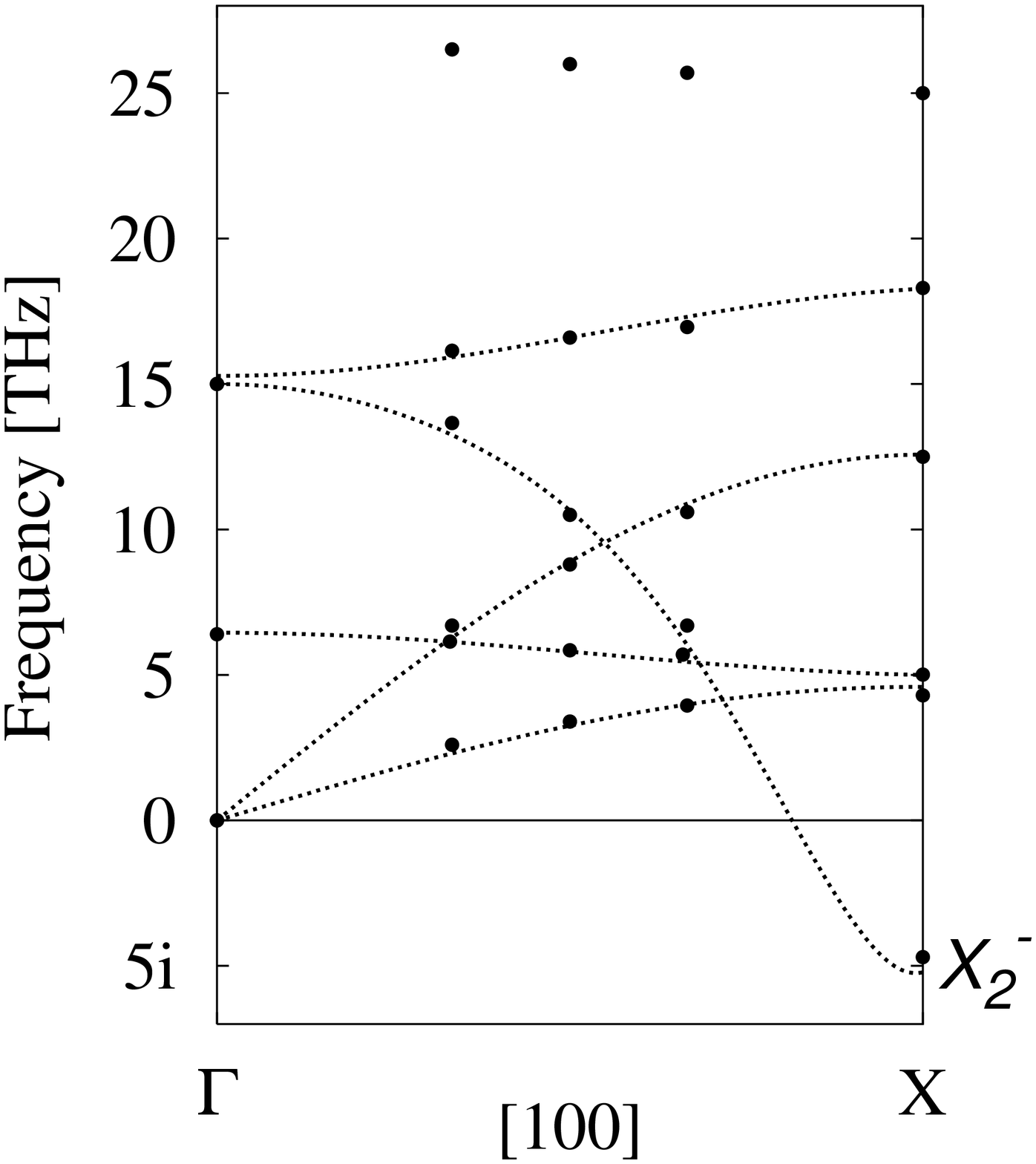,width=6cm,angle=0}} 
\end{figure}

Our LDA and TB results for the depth of the double well at the $t$ phase
equilibrium volume, V$_2$, are consistent with the recent LDA values of
$\approx$ 7 mRy.~\cite{Stapper99,Jomard99} This energy barrier for the
6-atom unit cell corresponds to a temperature of $\approx$ 1100 K.  The
same result was obtained by Jansen \cite{Jansen88} with the FLAPW method
who proposed a value of $\approx$ 1200 K. It is natural that these
temperatures, extrapolated from the 0 K potential energy, underestimate
the experimental phase transition temperature of 2570 K.~\cite{Aldebert85} The
experimentally observed phase transition temperature can be considered as the sum of
the kinetic contributions of all the activated eigenmodes, while the
calculated energy barrier refers to the kinetic contribution of the
$X_2^-$ eigenmode only. Even though it is reasonable to expect that at
the phase transition the soft mode in the phonon spectra (Figure \ref{phon}) will be
highly weighted in the total density of states, the kinetic energy $kT$
associated with all the other modes of vibrations will still contribute
to the measured phase transition temperature.

\subsubsection{Physical interpretation of the double well}

 What causes the $c \leftrightarrow t$ symmetry breaking? The tetragonal
 distortion of the oxygen sublattice implies the following geometrical
 changes: (i) Two Zr-O bond lengths get smaller and two get longer but
 the average Zr-O distance increases. (ii) entire columns of oxygen atoms
 shift one with respect to each other (see Fig.~\ref{cellct}) therefore
 the nearest neighbor O-O distances along the column remain constant
 while the other 4 nearest neighbor O-O distances increase. (iii) All
 the Zr-Zr distances remain constant. The overall increase of both the
 Zr-O and the O-O bond lengths is the basis of our interpretation of the
 double well, founded mainly on electrostatic arguments.

 By adjusting the various parameters describing ionicity, covalency and
 polarizability of the TB model we can select and isolate the effects
 that induce the double well, but before doing so it is instructive to understand
 how a simple RIM answers to the same question. It has been shown
 \cite{Wilson96zirc} that it is possible to reproduce the double well with a RIM
 in which there are two contributions: a repulsive short ranged pairwise
 interaction $U^{\rm pair}$ and a long ranged electrostatic term
 $U^{ii}$.

\begin{equation} U^{\rm RIM} = \sum_{i<j} A \, e^{ - \, b \, r_{ij} }
\, + \, \sum_{i<j}
\frac{z_i \, z_j}{r_{ij}} = U^{\rm pair} + U^{ii}, \label{RIM}
\end{equation}
$z$ is the ionic charge and $r_{ij}$ is the interatomic distance
between the ions $i$ and $j$. 

\begin{figure}
\caption{$\delta$ dependence of: (a) Madelung potential, (b) self-consistent charge
$Q=Q^e+Q^i$, (c) Electrostatic and Hubbard energies as in
Eq.~(\ref{elcst-en}), (d) the same including dipoles and quadrupoles. The
zero of energy is the top of the double well at V$_2$, total energies are in
Ry/formula unit, other quantities in a.u./ion.}
\label{split1}
\centerline{\psfig{file=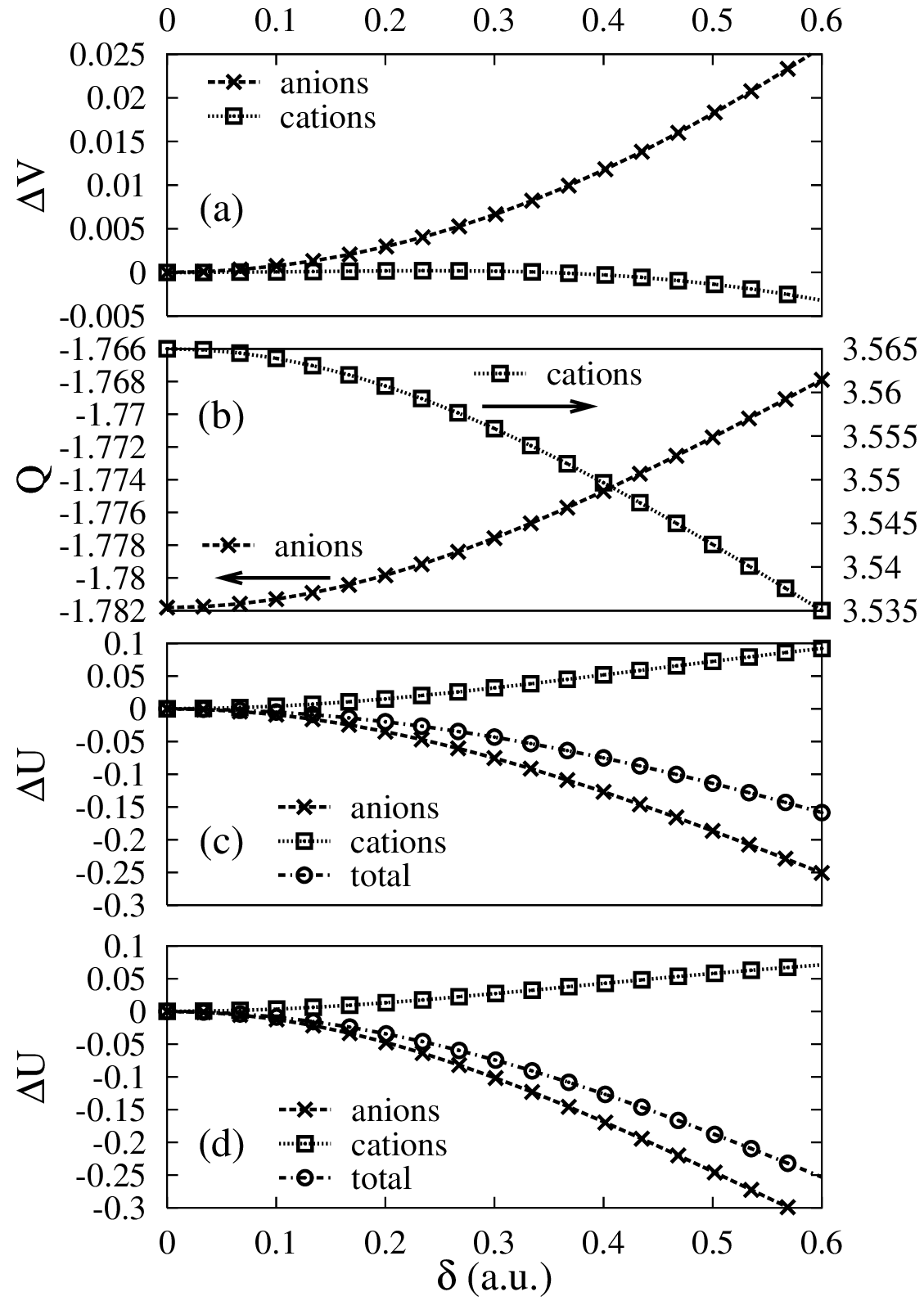,width=8.5cm,angle=0} }
\end{figure}

 The Zr-O bonds increase and decrease in length in a symmetric way.  As
 a net result, the centrosymmetric position of the oxygen atoms is a
 relative maximum of the Coulomb energy $U^{ii}$. The change in the
 Madelung potential caused by the tetragonal distortion is shown in
 Figure \ref{split1} (a). The overall increase of the O-Zr and O-O
 distances makes the oxygen sites much more sensitive to the change of
 the Madelung potential then the zirconium ones. The structural
 instability can therefore be interpreted as an effective way of
 minimizing the electrostatic energy of the oxygen sublattice. The
 repulsive Zr-O interaction counteracts the structural instability
 driven by the electrostatics, in a way which dominates at large
 displacements because of the exponential distance dependence of this
 repulsion. The double well shape of the energy profile is due to the
 different functional form of these opposing energy terms of
 Eq.(\ref{RIM}). This argument clearly depends on the strength of the
 repulsion, and does not work if the repulsion is too weak.

 It can be noticed that analogous terms are present in the TB model and a
 similar interpretation is tempting. However, we now have the additional
 effects due to polarization, covalency, and charge
 redistribution. Figure \ref{split1} (b) shows that the absolute value of
 the self-consistent equilibrium charge $Q$ decreases on both
 species. Consequently, in this approximation, the on-site energy
\begin{equation}
\frac{1}{2} \sum_{{\bf R}L} \left[ U \, \delta N^2_{{\bf R}\ell} +
               Q_{{\bf R}L} \, V_{{\bf R}L} \right] \, , \label{elcst-en}
\end{equation}
plotted in Figure \ref{split1} (c), decreases not only because of the
 previous geometric arguments but also because the charge redistribution
 reduces the ionic charges and therefore both the O-O and Zr-Zr
 electrostatic interactions.

 It is interesting to note that, on the oxygen atoms, the self-consistent
 charge $|Q^e|$ decreases with $\delta$ even though the total on-site
 potential [the sum of the Hubbard and electrostatic terms as in
 Eq.(\ref{elcst-en})] increases. This non-intuitive behavior of the
 charge transfer is due to covalency. The charge transfer is controlled
 both by the on-site potential and by the bonding integrals, which depend
 on the Zr-O distance. For $\delta \ne 0$, the overall increase in the
 Zr-O distance results in a decrease in the magnitude of the hopping
 integrals, and this overcomes the opposing effect of change in the
 on-site potential, {\em pushing back} some electrons from the oxygen to
 the zirconium sites.

 In the CIM-DQ, it was the quadrupole polarization of the O ions which
 stabilized the tetragonal structure, so it is of interest to see if it
 is also the development of a quadrupole moment in the tetragonal phase
 which stabilizes it within the SC-TB model.

\begin{figure}
\caption{Double well in the TB total energy at V$_2$ : ({\boldmath
$\times$}) no coupling between the potential and the oxygen atomic
orbitals; ({\boldmath $\circ$}) with dipoles and quadrupoles on the
oxygen atoms.}
\label{split2}
\centerline{\psfig{file=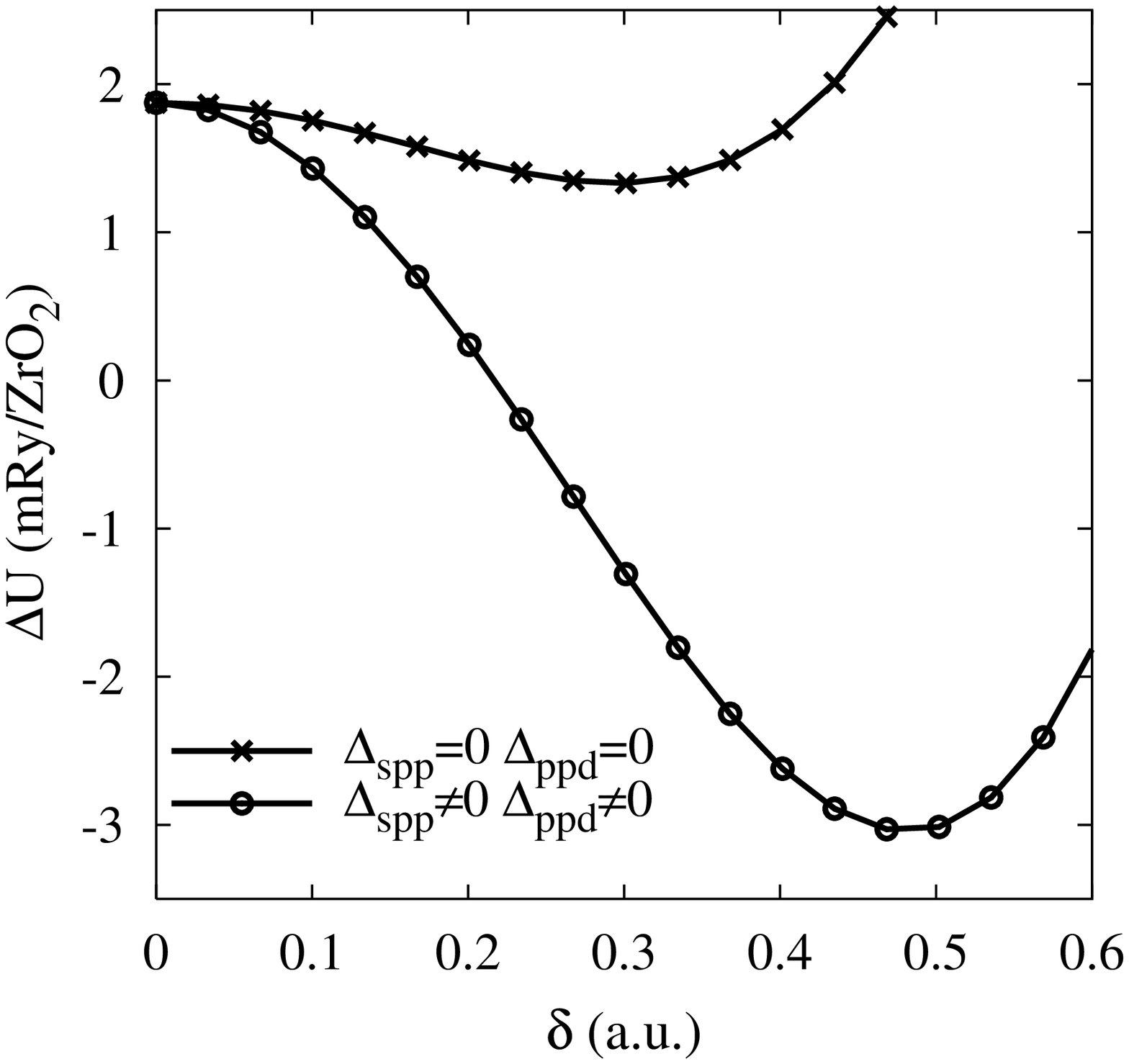,width=8.5cm,angle=-90}} 
\end{figure}

 In fact it turns out that covalency is the main effect, although
 polarizability is still significant. The $t$ structure is stable with
 respect to the $c$ one even with a {\it non} polarizable SC-TB model
 [Figure~\ref{split2} (a)]: the small energy difference is due to both
 ionicity and covalency of the crystal. The addition of the oxygen
 polarizability enhances the energy difference between the two phases
 deepening and broadening the double well [Figure~\ref{split2} (b)].

 We can be more specific about the nature of the polarization. In the $c$
 structure, the first non-zero components of the electrostatic potential
 are $V_0$ and $V_3$. The latter could, in principle, induce an octapole
 moment $Q_3$ on the anions. We truncated the multipolar expansion of the
 atomic multipole moments to the quadrupoles $Q_2$ therefore, within this
 approximation, the ions in the $c$ structure are not polarized. Higher
 order terms can be included in the expansion, but the overall agreement
 of the results with both experiments and first-principle calculations
 demonstrates that the model is already capturing the important physics
 of the system.

 As the anion sublattice is distorted, the symmetry lowering induces the
 $\ell=1$ and $\ell=2$ components of the potential which couple the $s$
 and $p$ oxygen atomic orbitals. The magnitude of the coupling, and
 therefore of the multipole moments, is controlled by the parameters
 $\Delta_{spp}$ and $\Delta_{ppd}$. The latter, fixed in order to
 reproduce the electronic structure of the rutile phase, produces very
 weak quadrupole moments, whose contribution to the double well is
 negligible. The former controls the size of the dipole moments whose
 symmetric distribution further minimize the electrostatic energy [Figure
 \ref{split1} (d)]. The total effect on the double well is shown in
 Figure~\ref{split2}.

\subsubsection{Landau theory} 

The $c \leftrightarrow t$ phase transition can be interpreted in terms of
the Landau Theory.~\cite{Landau5} In a subsequent paper we plan to
explore the free energy surface at $T>0$ with this formalism, so it is
convenient to introduce it here to discuss the $T=0$
results. Experimentally, the mechanism of this phase transition has been
very controversial and a clear description is still
missing.~\cite{Cohen63,Sakuma85,Heuer84,Heuer87,Stubican78,Sakuma87,Zhou91,Katamura97}

Theoretically, Chan \cite{Chan88} suggested that a partial softening of
an elastic constant is the driving force of this phase transition and,
after symmetry considerations based on the elastic strains only,
concluded that the phase transition must be of first order.  We show here
that the inclusion of the order parameter $\delta$ gives a second order
phase transition.  A similar discussion has been given by Ishibashi and
Dvo\'{r}\u{a}k.~\cite{Ishibashi89}

According to the Landau Theory, the appropriate thermodynamic potential
which describes the relationship between the two phases of interest, is
expanded in a Taylor series in one or more order parameters, in which the
expansion coefficients are temperature dependent.  The order parameters
are non-zero in the low symmetry phase and vanish in the high symmetry
one, providing therefore a unique way to differentiate the two phases.
The terms involved in the Taylor expansion are invariants under the
symmetry operations of the high symmetry phase and can be identified
using group theory.

In the case of zirconia, the $c$ structure is unstable along the three
crystallographic directions, therefore the distortions along $x,y$ and
$z$ have to be explicitly treated in the energy expansion. This suggests
the following 9 order parameters, defined in terms of the strain tensor
{\boldmath $ \epsilon $}, and grouped into 4 symmetry-adapted bases which
spans the corresponding irreducible representations:

\begin{eqnarray*}
\delta_x , \delta_y , \delta_z   & \hspace{1cm}  & {\rm T_1} \\
\epsilon_{xx} + \epsilon_{yy} + \epsilon_{zz}  &  &{\rm A_1} \\
\left(2\epsilon_{zz} - \epsilon_{xx} - \epsilon_{yy}
\right), \, \sqrt{3} \left(\epsilon_{xx} - \epsilon_{yy} \right) 
& & {\rm E} \\ 
\epsilon_{xy}, \epsilon_{yz}, \epsilon_{zx} & & {\rm T_2}
\end{eqnarray*}

A complete analysis involving all the order parameters will be done in a
separate paper, here we simplify the total energy expansion selecting one
of the three possible directions of the tetragonal axis. Under this
hypothesis three order parameters are necessary to describe the $c
\leftrightarrow t$ phase transition of zirconia: $\delta$, $\eta$ and
$\eta_0$. The high temperature $c$ phase has the full cubic symmetry
$m3m$ and the only degree of freedom is the hydrostatic strain $\eta_0=
\epsilon_{xx} + \epsilon_{yy} + \epsilon_{zz}$. The low-symmetry $t$
phase is defined by the distortion of the anionic sublattice $\delta$,
which we define as the amplitude of the $X_2^-$ mode of vibration, and by
the tetragonal strain $\eta=\left(2\epsilon_{zz} - \epsilon_{xx} -
\epsilon_{yy} \right)$.

The three order parameters can be hierarchically classified according to
the amount of symmetry breaking that they involve. The hydrostatic strain
$\eta_0$ preserves the cubic symmetry of the crystal. The
tetragonal strain $\eta$ maintains the number of atoms in the primitive
cell and lowers the symmetry to the point group $4/mmm$ which still has
the mirror symmetry operation perpendicular to the tetragonal axis. The
tetragonal distortion $\delta$ breaks this symmetry operation and
involves cell doubling. Therefore, according to Landau theory, $\delta$
is the primary order parameter, $\eta$ is the secondary and $\eta_0$ is
the tertiary one.

\begin{figure}
\caption{Energy-Volume curves for the $c$ and $t$ structures:
note the convergence at small volumes V$_1$. V$_0$ and V$_2$ are
the equilibrium volumes of the $c$ and $t$ phases respectively.}
\label{envol-ct}
\centerline{\psfig{file=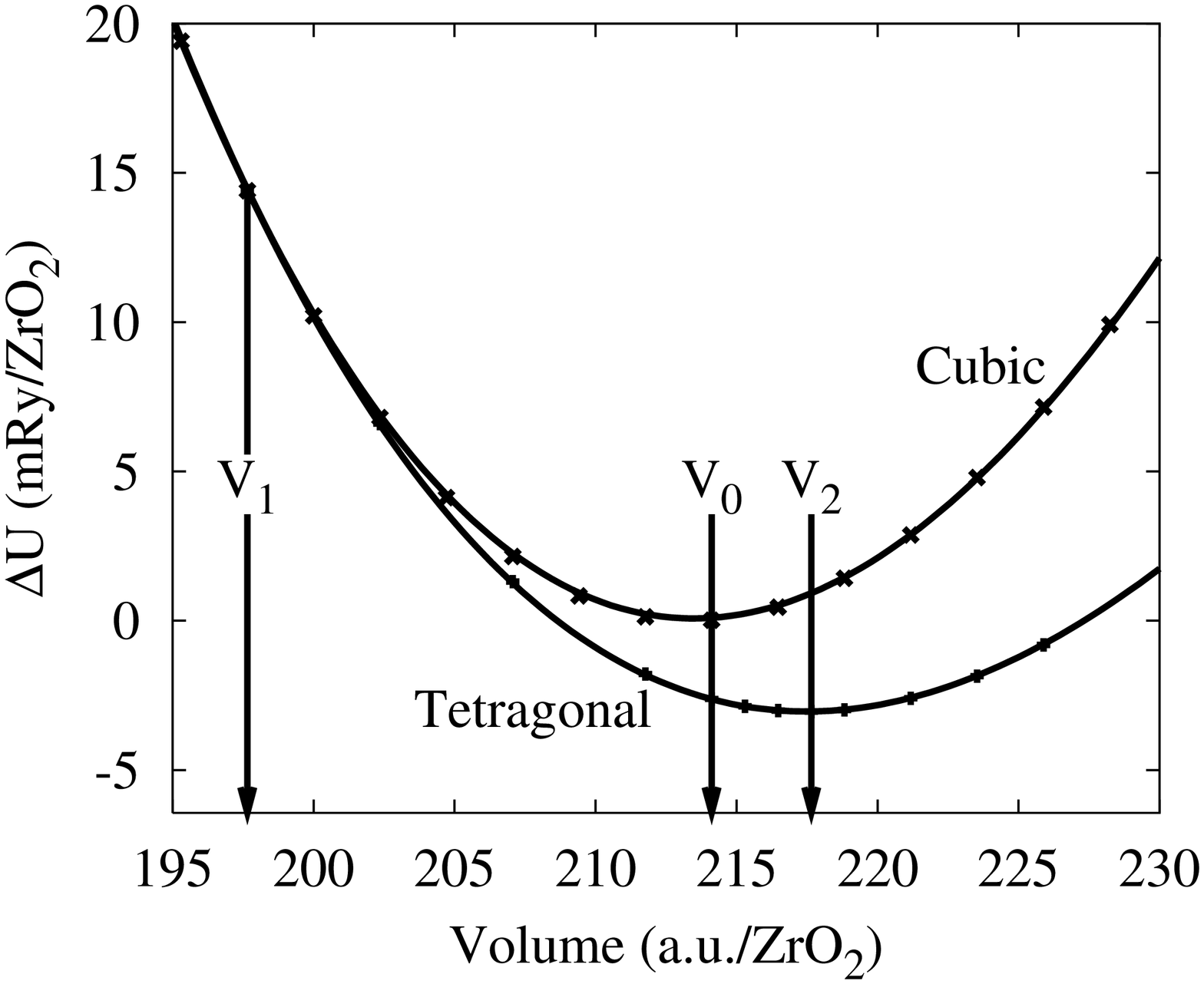,width=8cm,angle=-90}} 
\end{figure}

The potential energy is expanded as a power series in these order
parameters around the equilibrium volume of the cubic phase $V_0$
(Figure \ref{envol-ct}):

\begin{eqnarray} \label{lanexp}
U - U^{c}_{V_0} & = &  \frac{a_2}{2} \: \delta^2 +
\frac{a_4}{4} \: \delta^4 + b_0 \: \delta^2 \eta_0 + b_1 \: \delta^2
\eta + \\ \nonumber 
& & \frac{c_0}{2} \: \eta_0^2 + \frac{c_1}{2} \: \eta^2 + {\cal O}
(\delta^6).   
\end{eqnarray}
The elastic constants $c_0$ and $c_1$ are proportional respectively to
the bulk modulus and to $C^{\prime}=\frac{1}{2}(c_{11}-c_{12})$ in the
$c$ phase described in the next section.  The third order term
$\delta^3$ is forbidden by symmetry, therefore this transition is of
second order if $a_2$ goes negative.

\begin{figure}
\caption{Volume dependence of the order parameters calculated with the TB
model: $\eta_0$ is the hydrostatic strain of the cubic cell from the
reference volume V$_0$, $\eta$ is the tetragonal strain of the cell,
and $\delta$ (a.u.) is the tetragonal distortion of the oxygen sublattice.}
\label{landvol}
\centerline{\psfig{file=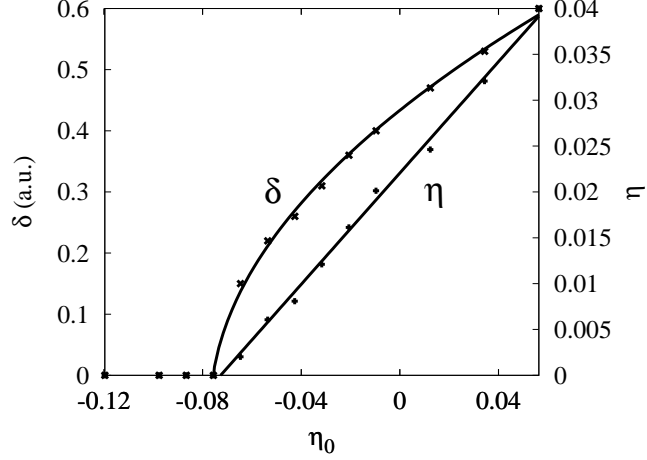,width=8.5cm,angle=-90} }
\end{figure}

The volume dependence of the order parameters can be studied by setting
to zero $\nabla_\eta U$ and $\nabla_{\eta_0} U$. Both the {\it ab initio}
and TB results (Figure~\ref{landvol}) confirm the analytic expressions:

\begin{eqnarray}
\left\{ \begin{array}{lcl}
\eta & = & - \frac{b_1}{c_1}\, \delta^2 \\
& & \\
\eta_0 & = & - \frac{b_0}{c_0} \, \delta^2 
\end{array} \right. &
\hspace{0.5cm} \Rightarrow \hspace{0.5cm} & 
\left\{ \begin{array}{lcl}
\delta & \propto &  \sqrt{\eta_0} \\
& & \\
\eta   & \propto & \eta_0
\end{array} 
\right. \label{land-orvol}
\end{eqnarray}

These expressions show that the second-order strain terms of
Eq.(\ref{lanexp}) are already proportional to $\delta^4$ and therefore,
within the chosen order of approximation, it is not necessary to
include third-order terms in $\epsilon_{ij}$. Moreover, from the static
results it is clear that the description of the high temperature
stability of the $c$ phase must go beyond the quasi-harmonic
approximation. The higher the temperature, the larger the volume and,
according to Figure~\ref{landvol}, the larger $\delta$ and $\eta$.
Therefore, in a simple quasi-harmonic picture, a higher temperature
seems to favor the $t$ phase with respect to $c$, in contradiction to
the experimental observation.

\begin{figure}
\caption{SC-TB total energy versus tetragonal distortion $\delta$. (a) Fit
of the data with the Landau energy expansion Eq.(\ref{lanexp}); (b)
transferability of the coefficients at values of hydrostatic ($\eta_0$)
and tetragonal ($\eta$) strains different from the reference ones.}
\label{lansurf}
\centerline{\psfig{file=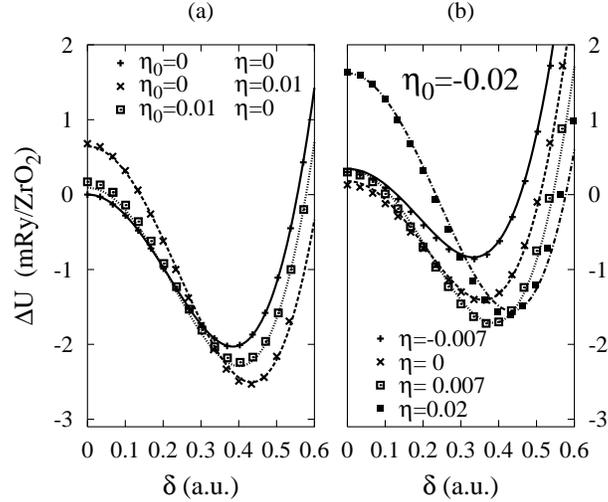,width=8.5cm,angle=-90}} 
\end{figure}

The parameters $c_1$ and $c_0$ are known from the elastic properties of
the crystal and have been calculated independently (see next section).
The coefficients $a_2$ and $a_4$ have been fitted to the double well of an
undistorted stress-free cubic crystal (in the sense $\eta=0$ and $\eta_0=0$).  In a
similar way, $b_1$ and $b_0$ have been fitted to the double well of a tetragonal
crystal at $V_0$ ($\eta_0=0$, $\eta \ne 0$) and of a cubic crystal near
$V_0$ ($\eta=0$, $\eta_0 \ne 0$) respectively. Figure~\ref{lansurf} (a)
shows the three curves used for the fitting procedure.  The agreement is
very good even far away from the reference volume of the energy
expansion [Figure~\ref{lansurf} (b)]. This demonstrate that the fourth
order truncation in Eq.(\ref{lanexp}) is sufficient to capture all the
essential features of the 0 K energy surface.  

Nardelli {\it et al.} \cite{Nardelli92,Nardelli95} have shown the crucial
r$\hat{\rm o}$le played by the coupling between different order
parameters and how it can affect the correct interpretation of the phase
transformation. To see this we substitute the relationships
(\ref{land-orvol}) back in Eq.(\ref{lanexp}):

\begin{equation}
U - U^{c}_{V_0} = \frac{a_2}{2} \delta^2 + \left[ \frac{a_4}{4} -
\frac{b_0^2}{2 \, c_0}  - \frac{b_1^2}{2 \, c_1} \right] \delta^4 + {\cal O}\left(
\delta^6 \right).
\end{equation}
The above equation shows that the coupling term $\left( \frac{b_0^2}{2 \,
c_0} + \frac{b_1^2}{2 \, c_1} \right)$ can renormalize the fourth order
coefficient, and could make it negative. In that case it would be
necessary to truncate Eq.(\ref{lanexp}) at the sixth order term in
$\delta$, including therefore the third-order terms in the strain. These
would then drive the phase transition making it first order.~\cite{Chan88,Anderson65}
The numerical values of the coefficients (Table~\ref{landcfc}) allow us
to estimate the amount of the coupling. We find that the coupling term is
$\approx 20\%$ of $\frac{a_4}{4}$, not big enough to affect the sign of
the fourth order coefficient and therefore the 0 K calculations suggest
that the phase transition is displacive of second order.

The temperature dependence of the elastic constants might change this
description and the final answer will be given by high temperature MD
calculations which are in progress. 

\section{Distortions}
\label{distortions}

\subsection{Elastic constants}

The elasticity of $c$ and $t$ zirconia has been explored with the TB
model. The analysis involved the distortion of the crystal along high
symmetry directions, the calculation of the total energy for different
values of the distortion parameter and the fit of the results to a
polynomial.  The rigidity of the crystal with respect the particular
distortion applied has been extracted from the quadratic coefficient of
the energy series expansion. For each strain of the $t$ structure, we
constrained the volume to the predicted equilibrium value and minimized
the energy with respect to the internal degrees of freedom.

Volume conserving stretches along the high symmetry directions of the $c$
unit cell $<\!\!100\!\!>$ and $<\!\!111\!\!>$ provide
$C^\prime=\frac{1}{2}(c_{11}-c_{12})$ and $c_{44}$ respectively. Extra
distortions are necessary when the symmetry is lower: if $z$ is the
tetragonal axis, an independent set of 5 shear moduli were obtained by
stretching along $<\!\!100\!\!>$, $<\!\!001\!\!>$, $<\!\!111\!\!>$,
$<\!\!110\!\!>$ and $<\!\!101\!\!  >$. The bulk moduli have been obtained
by fitting the Energy-Volume curves with a Birch-Murnaghan Equation of
State.~\cite{Murnaghan44,Birch78}

Liu {\it et al.}~\cite{Liu87}used the slope of the acoustic branches
at small wavelength of a ZrO$_2$-Y$_2$O$_3$ (15 $\%$) system to
estimate the elastic constants of the cubic phase. Kandil {\it et
al.}~\cite{Kandil82} directly measured the elastic constants of Yttria
Stabilized Zirconia (YSZ) single crystals: the reference values included
in Table~\ref{elconst} are extrapolations to 0$\%$ impurities. To our
knowledge there is no equivalent experimental study of the elasticity
of the $t$ phase. The most recent values \cite{Kisi98b} are measured via
a powder diffraction technique on 12\% Ce-doped $t$ zirconia. 


We compare our predictions with theoretical and experimental data in
Table~\ref{elconst}. The results of two other theoretical approaches, the
Hartree-Fock and the PIB ones, are very different. As already
mentioned in the Introduction, none of these calculations predicted the
correct relative energetics of the crystal structures. Elasticity is a
property of the energy second derivative: a good description of the
energy curves is a prerequisite for reliable elastic constant
calculations.

The fairly good agreement of our calculations with the experiments
further indicates that the SC-TB model captures the main physics of the
bonding. The bulk modulus, however, is seriously overestimated: this
may not be an intrinsic limitation of the TB model, because it was fit
precisely to the NFP-LMTO calculation, which similarly overestimates
this quantity.

\subsection{Phonon Spectra}

In order to test the model further, as well as to give further insight
into the spontaneous symmetry breaking of the $c$ phase, we studied its
vibrational properties. First principle calculations
\cite{Detraux98,Parlinski97} predict an imaginary frequency at the
boundary of the BZ: this reinforces the idea that the phase transition is
displacive, and driven by the softening of an optic mode.

Our calculations were carried out with the TB model on a 96-atom
unit cell. The eigenvalues and eigenvectors of
the possible vibrational modes in that unit cell, were found by
diagonalising the dynamical matrix which we calculated using the direct
method. The procedure was as follows.

Within the harmonic approximation, the potential energy $\Phi$ is
expanded to second order in powers of the atomic displacements ${\bf
u}$:

\begin{equation}
\Phi= \Phi_0 + \frac{1}{2}
\sum_{ \scriptsize \begin{array}{c} l,  \kappa, \alpha \\
		     l^{\prime} \! \! ,  \kappa^{\prime} \! \! ,  \beta
    \end{array}}  \Phi_{\alpha\beta} \left(  \! \! \begin{array}{c} {l l^\prime} \\ \kappa \kappa^\prime
\end{array}  \! \! \! \right) u_\alpha
\left( \! \!  \begin{array}{c} {l} \\ \kappa \end{array}  \! \!
\right) u_\beta \left( \! \! 
\begin{array}{c} {l^\prime} \\ \kappa^\prime \end{array}  \! \! \! \right) + \ldots 
\end{equation}
We use the notation of Maradudin {\it et al.}~\cite{Maradudin63}:
$\kappa$ and $l$ label respectively the atom in the primitive cell
and the position of the primitive cell with respect to some origin.  The
direct method consists in computing the force constants
$\Phi_{\alpha\beta}$ via total energy and force calculations. In general,
the atom  $\kappa$ in the $l$ cell is displaced by a small amount
in direction $\alpha$ and the Hellmann-Feynman forces on the other atoms
are recorded. These give directly the quadratic terms in the total
energy expansion.  The force constants $\Phi_{\alpha\beta}$ 
can be related to the
corresponding term of the dynamical matrix {\bf D} via the usual
relation:

\begin{equation} 
D_{\alpha\beta} \left(  \! \! \begin{array}{c} {\bf k} \\ \kappa \kappa^\prime
\end{array} \! \! \! \right) = \frac{1}{\sqrt{M_\kappa
M_{\kappa^\prime}}} \sum_l \Phi_{\alpha\beta} \left( \! \! \begin{array}{c} {l} \\ \kappa \kappa^\prime
\end{array}  \! \! \! \right) e^{-2 \pi {\bf k \cdot x{\it (l)}}} .
\end{equation}
$M_\kappa$ is the mass of the atom $\kappa$ and $ \bf k$ is a point in
the BZ. The crystal symmetry can considerably reduce the number of
necessary independent calculations.~\cite{Maradudin68,Warren68}

The phonon spectra plotted along the high symmetry direction
$<\!\!100\!\!>$ are shown in Figure~\ref{phon}. The main feature of the
spectra is the imaginary frequency of the $X_2^-$ mode of vibration which
corresponds to the tetragonal instability shown in Figure~\ref{well}. As
already mentioned the tetragonal instability involves cell doubling
therefore the corresponding eigenvector appears at the BZ border of the
$c$ phase. The soft mode at the $X$ point $\nu_s=5.1i$ is the natural
consequence of the negative curvature of the energy surface at $\delta=0$
(Figure~\ref{well}). Setting to zero the dipolar polarizability of the
anions ($\Delta_{spp}=0$), the $X_2^-$ mode is still soft,
$\nu_s=0.8i$, but the force constant corresponding to the instability
is much smaller. This is consistent with Figure~\ref{split2} where the
same effect is studied from the energetic point of view: the energy curve
is concave at $\delta=0$ even when the oxygens are not polarizable.

The effect of the oxygen polarizability is evident on the $T_{1u}$
IR-active mode, which involves the rigid displacement of the two atomic
sublattices. The calculated vibration frequency is 7.9 THz when the
anions are not polarizable and 6.3 THz when the dipolar degree of freedom
is allowed. The closer agreement of the non-polarizable result with the
DFT frequencies of $8.1 - 8.5$ THz, together with the overestimation of
the bulk modulus suggests that the present model could slightly
overestimate both the short-range repulsion between closed shells of
electrons, responsible for the high bulk modulus, and the long range
polarization effects which make the $T_{1u}$ frequency lower than the
{\it ab initio} values. The results might be improved with a more
accurate re-parameterization but the physical interpretation of the {\it
ab-initio} results, which is the main objective of this analysis, is
unlikely to change.

Table~\ref{phontab} shows the general agreement of the TB model with
other calculations and with the experimental data. The latter are
measured by Raman spectroscopy and inelastic neutron scattering at high
temperatures on YSZ.

Certain non-analytical terms in the dynamical matrix have been neglected,
namely those relating to macroscopic polarization or the Berry phase.
For this reason our calculations cannot reproduce the LO-TO splitting of
12 THz calculated by Detraux {\it et al.}.~\cite{Detraux98} The
non-analytical terms can be approximated by knowing the Born effective
charge and the dielectric tensor, both of which could in principle be
obtained from our model. This has previously been done in a TB framework
,~\cite{Bennetto96} although not for ZrO$_2$, and we plan to investigate
the effect in the future.

\section{Conclusions}

We have explored the predictive power of a polarizable SC-TB model by
investigating the crystal stability of pure zirconia. The results of this
extended TB model are in overall good agreement with our own {\it ab
initio} (NFP-LMTO) calculations and with previous experimental and
theoretical {\it ab initio} studies. This semiempirical model has
captured the basic physics of the relative phase stability of zirconia
with a set of parameters which are transferable between the crystal
structures. A noteworthy improvement over all previous models is the
absolute stability of the monoclinic structure at 0 K with respect to the
usual set of alternatives.  This demonstrates that the model is ready to
deal with more complicated crystalline environments such as solid
solutions, high temperature distortions, or interfaces.

The TB model predicts that the covalent character of the Zr-O bond
 plays a major r$\hat{\rm o}$le in the energetics of zirconia, more so
than the polarizability of the oxygen ions. For example, the double
well about the cubic structure, absent in a rigid ion model, exists
when covalency is included; it is further enhanced by including also
polarizability at the dipole level. We do not believe that the
separation between covalent effects and polarizability effects is
unique, since it depends on the choice of basis functions. Quite
possibly the previous polarizable ion models were  capturing some
effects of charge redistribution which could alternatively be described
by covalency. It remains to be seen if a model for zirconia without
explicit covalency could satisfactorily reproduce all the structural
energies.

 The Landau Theory, used to interpret the TB and {\it ab initio} results,
together with the lattice dynamic analysis, shows that the $c
\leftrightarrow t$ phase transition is displacive of the second order and
is driven by the softening of the $X_2^-$ mode of vibration. If it had
been driven by a softening of the corresponding elastic constant
$c_{11}-c_{12}$ it would have been a first order transition. The partial
softening of the elastic constants due to the temperature could also in
principle change the character of the phase transition. We are currently
applying the molecular dynamics technique to understand the high
temperature thermodynamic stability of the $c$ phase and to explore the
character of the phase transition.  To this end we can use thermodynamic
integration to go beyond the harmonic approximation. The preliminary
results of these calculations will appear in the near
future.~\cite{MDzirc} 

Since the valence electrons are treated explicitly within the
SC-TB model we also hope to be able to study the effects of point
defects. This would be more difficult with a classical polarizable ion model
because of the problems associated with charge conservation and
redistribution.

\acknowledgements

SF is grateful for support from the European Science Foundation, Forbairt
and the British Council, and for discussions with John Corish and Nigel
Marks. ATP and MWF are grateful to the EPSRC for funding under Grants
No. L66908 and No. L08380. This work has been supported by the European
Communities HCM Network ``Electronic Structure Calculations of Materials
Properties and Processes for Industry and Basic Science'' under grant
No. ERBFMRXCT980178.



\begin{table}
\caption{Parameters of the polarizable SC-TB model. Energy in Ry and
lengths in atomic units.}
\label{param}
\begin{tabular}{cc}
\multicolumn{2}{c}{\it On site parameters} \\
 ${\cal H}^0_{s}$ = 0.35  & $U_s$ = 1  \\
 ${\cal H}^0_{p}$ =-0.70  & $U_p$ = 1  \\
 ${\cal H}^0_{d}$ =-0.10  & $U_d$ = 1  \\ 
\hline
\multicolumn{2}{c}{\it Bond integrals } \\
\multicolumn{2}{c}{$ V_{ll^\prime}
       \left( \frac{d}{r} \right)^n 
       \exp \left\{ n \left[ - \left( \frac{r}{r_c} \right)^{n_c} +
                      \left( \frac{d}{r_c} \right)^{n_c}
                      \right] \right\} $ } \\
\begin{tabular}{cccccc}
 & $V_{ll^\prime}$ & $n$  & $n_c$ & $d$ & $r_c$ \\ \hline
$ss\sigma$ & -0.060 & 2 & 0 & 4.90 & 6.24 \\ 
$sp\sigma$ & ~0.070 & 2 & 0 & 4.90 & 6.24 \\
$pp\sigma$ & ~0.050 & 3 & 4 & 4.90 & 6.24 \\
$pp\pi$    & -0.008 & 3 & 4 & 4.90 & 6.24 \\
$sd\sigma$ & -0.050 & 3 & 0 & 4.24 & 4.90 
\end{tabular} &
\begin{tabular}{cccccc}
 & $V_{ll^\prime}$ & $n$  & $n_c$ & $d$ & $r_c$ \\ \hline
$pd\sigma$ & -0.100 & 4 & 0 & 4.24 & 4.90  \\
$pd\pi$    & ~0.058 & 4 & 0 & 4.24 & 4.90  \\
$dd\sigma$ & -0.050 & 5 & 0 & 6.02 & 6.93  \\
$dd\pi$    & ~0.033 & 5 & 0 & 6.02 & 6.93  \\
$dd\delta$ & ~0.008 & 5 & 0 & 6.02 & 6.93  
\end{tabular} \\
\hline
\multicolumn{2}{c}{\it Polarization terms} \\
$\Delta_{spp}$ = 0.73 & $\Delta_{ddd}$ = 0~~~~ \\
$\Delta_{ppd}$ = 1.89  & $\Delta_{ddg}$ = 63.5 \\
\hline
\multicolumn{2}{c}{\it Pair potential} \\
\multicolumn{2}{c}{$ U(r)= A \, e^{\left(-b \, r \right)} $} \\
$A$ = 181.972  & $b$ = 1.652 \\
\end{tabular}
\end{table}

\begin{table*}
\caption{Equilibrium structural parameters for the 0-pressure phases of
ZrO$_2$. The lattice parameters $a$, $b$, $c$ (a.u.), and the volumes
(a.u./ZrO$_2$) of the $c$, $t$, and $m$ structure are referred to the
12-atoms, 6-atoms, and 12-atoms unit cells respectively. $\delta$ denotes
the internal degree of freedom of the $t$ phase (see Fig.~\ref{cellct}),
$\beta$ is the angle of the $m$ cell in degrees, and $x$, $y$, $z$ are
the fractional coordinates of the non-equivalent sites in the $m$
structure.}
\label{strucpar}
\begin{tabular}{c|cccccc}
       & Expt.\tablenote[1]{The experimental values of the cubic and
       tetragonal structures have been extrapolated at 0 K using the
       thermal expansion data from Ref.~\onlinecite{Aldebert85}}   
                  & SC-TB      & NFP-LMTO     & PW-PP & PW-PP & FLAPW \\ 
       & Refs.~\onlinecite{Aldebert85,Howard88} & this work & this work & Ref.~\onlinecite{Kralik98} &
Ref.~\onlinecite{Stapper99} & Ref.~\onlinecite{Jansen91} \\ \hline
 & \multicolumn{6}{c}{$Cubic$} \\
{Volume }  & 222.50 & 213.40 & 210.33 & 215.29 & 220.84 & 217.81 \\ 
$a$        & 9.619  & 9.486  & 9.442  & 9.514  & 9.595  & 9.551 \\ \hline
 & \multicolumn{6}{c}{$Tetragonal$} \\
{Volume }  & 229.93 & 217.73 & 215.16 & 218.69 & 225.31 & 218.84 \\ 
$a$        & 6.748  & 6.709  & 6.695  & 6.734  & 6.797  & 6.747  \\ 
$c/a$      & 1.451  & 1.442  & 1.434  & 1.432  & 1.435  & 1.425  \\ 
$\delta/c$ & 0.057\tablenote[2]{At 1568 K}  
                    & 0.047  & 0.051  & 0.042  & 0.042  & 0.029  \\ \hline
 & \multicolumn{6}{c}{$Monoclinic$} \\
{Volume } & 237.67 & 222.89 & 226.13 & 230.51 & 236.46 & \\ 
$a$       & 9.733  & 9.592  & 9.417  & 9.611  & 9.733\tablenote[3]{Fixed
       to the experimental values of Ref.~\onlinecite{Howard88}} & \\ 
$b/a$   & 1.012  & 1.001  & 1.036 & 1.024  & 1.012$^{\rm c}$ & \\ 
$c/a$   & 1.032  & 1.019  & 1.057 & 1.028  & 1.032$^{\rm c}$  & \\ 
$\beta$ & 99.23  & 98.00  & 98.57 & 99.21  & 99.23$^{\rm c}$  & \\ 
 & & &  & & & \\ 
$x_{\rm Zr}$  & 0.275 & 0.272 & 0.274 & 0.278 & 0.277 & \\ 
$y_{\rm Zr}$  & 0.040 & 0.027 & 0.040 & 0.042 & 0.043 & \\ 
$z_{\rm Zr}$  & 0.208 & 0.217 & 0.212 & 0.210 & 0.210 & \\ 
$x_{\rm O_1}$ & 0.070 & 0.078 & 0.069 & 0.077 & 0.064 & \\ 
$y_{\rm O_1}$ & 0.332 & 0.336 & 0.339 & 0.349 & 0.324 & \\ 
$z_{\rm O_1}$ & 0.345 & 0.342 & 0.338 & 0.331 & 0.352 & \\ 
$x_{\rm O_2}$ & 0.450 & 0.452 & 0.448 & 0.447 & 0.450 & \\ 
$y_{\rm O_2}$ & 0.757 & 0.752 & 0.753 & 0.759 & 0.756 & \\ 
$z_{\rm O_2}$ & 0.479 & 0.472 & 0.478 & 0.483 & 0.479 & \\ 
\end{tabular}
\end{table*}

\begin{table}[t]
\caption{Energy differences (mRy/ZrO$_2$) between the zirconia polymorphs
and the $c$ phase calculated at the minimized structural parameters of
Tables~\ref{strucpar} and~\ref{parort}. The experimental values are
derived from enthalpy differences at the phase transition temperature.}
\label{energy}
\begin{tabular}{ll|cc|cc}
    &  & $\Delta U^{t-c}$ & $\Delta U^{m-c}$ & $\Delta U^{O_I-c}$ &
$\Delta U^{O_{II}-c}$ \\ \hline
Expt.   & Ref.~\onlinecite{Ackermann75}& -4.2 & -8.8  & & \\
SC-TB                                  &  & -3.0 & -7.4  & -3.6 & 2.8 \\
NFP                                 &  & -3.6 & -7.7  & - & \\
PW-PP   & Ref.~\onlinecite{Kralik98}   & -3.3 & -7.5  & - & \\
PW-PP   & Ref.~\onlinecite{Stapper99}  & -3.5 & -8.2  & -5.3 & - \\
PW-PP   & Ref.~\onlinecite{Jomard99}   & -1.5\tablenote[1]{LDA calculation}
                                         -5.9\tablenote[2]{Perdew-Wang GGC calculation}  
                                               & -4.4$^{\rm a}$ 
                                                 -13.9$^{\rm b}$
                                               & - & 0.7$^{\rm a}$
                                                     7.3$^{\rm b}$
\end{tabular}
\end{table}

\begin{table}
\caption{External and internal degrees of freedom of the orthorhombic
structures. Lattice parameters $a$, $b$, $c$ in a.u., volumes in
a.u./ZrO$_2$. The fractional coordinates of the non-equivalent sites are
denoted with $x$, $y$, and $z$.}
\label{parort}
\begin{tabular}{c|ccc|ccc}
 & \multicolumn{3}{c}{$Ortho \, I$} & \multicolumn{3}{c}{$Ortho \, II$} \\
        & Expt.    & SC-TB     & PW-PP  & Expt.   & SC-TB     & PW-PP \\ 
 & Ref.~\onlinecite{Ohtaka90proc} 
                & this work & Ref.~\onlinecite{Stapper99} 
                                     & Ref.~\onlinecite{Haines97} 
                                               & this work 
                                     & Ref.~\onlinecite{Jomard99} \\ \hline
Vol.    & 228.159 & 218.69 &  226.7 & 203.54 & 196.08 & 212.44 \\
$a$     & 19.060  & 18.737 &  19.060\tablenote[1]{Fixed to the experimental
 values of Ref.\onlinecite{Ohtaka90proc}} & 10.558 & 10.541 & 10.721 \\
$b/a$   & 0.522   & 0.520  &  0.522$^{\rm a}$ & 0.596 & 0.592 & 0.593 \\
$c/a$   & 0.505   & 0.511  &  0.505$^{\rm a}$ & 1.161 & 1.139 & 1.163 \\ \hline
$x_{\rm Zr}$   & 0.884 & 0.880 & 0.884 & 0.246 & 0.255 & 0.253 \\
$y_{\rm Zr}$   & 0.033 & 0.002 & 0.036 & 0.250 & 0.250 & 0.250 \\
$z_{\rm Zr}$   & 0.256 & 0.256 & 0.253 & 0.110 & 0.099 & 0.111 \\
$x_{\rm O_1}$  & 0.978 & 0.978 & 0.978 & 0.360 & 0.354 & 0.360 \\
$y_{\rm O_1}$  & 0.748 & 0.745 & 0.739 & 0.250 & 0.251 & 0.250 \\
$z_{\rm O_1}$  & 0.495 & 0.509 & 0.499 & 0.424 & 0.421 & 0.425 \\
$x_{\rm O_2}$  & 0.791 & 0.784 & 0.790 & 0.025 & 0.022 & 0.023 \\
$y_{\rm O_2}$  & 0.371 & 0.371 & 0.374 & 0.750 & 0.749 & 0.750 \\
$z_{\rm O_2}$  & 0.131 & 0.130 & 0.127 & 0.339 & 0.338 & 0.340 
\end{tabular}
\end{table}

\begin{table}
\caption{Coefficients (a.u.) of the energy Taylor expansion (\ref{lanexp}).}
\label{landcfc}
\begin{tabular}{clllc}
\hspace{0.5cm} & $a_2$ = -0.053 & $b_0$   = -0.062 & $c_0$   = 0.621 &
\hspace{0.5cm} \\ 
               & $a_4$ = ~0.347   & $b_1$ = -0.152   & $c_1$ = 0.818 & 
\end{tabular}
\end{table}

\begin{table}
\caption{Elastic constants (GPa) of the $c$ and $t$ structures.}
\label{elconst}
\begin{tabular}{c||ccccc}
 &  SC-TB    & Expt.    & PIB & HF    & DFT   \\
 & this work &  Refs.~\onlinecite{Liu87},~\onlinecite{Kandil82},~\onlinecite{Kisi98b}&
 Ref.~\onlinecite{Cohen88}& Ref.~\onlinecite{Orlando92} & 
 Ref.~\onlinecite{Stapper99} \\ \hline
 & \multicolumn{5}{c}{\it Cubic } \\
 K$_0$       & 310   & 194  254 & 288 & 222 & 268   \\
 $C^\prime$ & 175 & 167  165 & 195 & 304 &  -    \\
 $c_{44}$ & 57   & 47  61 & 180 & 82  &  -    \\ \hline
 & \multicolumn{5}{c}{\it Tetragonal} \\
 K$_0$    & 190  & 151  & 179  &  &  197  \\
 $c_{11}$ & 366  & 327 & 465 & - & -  \\
 $c_{33}$ & 286  & 264 & 326 & - & -   \\
 $c_{12}$ & 180  & 100 & 83  & - & -   \\
 $c_{13}$ &  80  &  62 & 49  & - & -   \\
 $c_{44}$ &  78  &  59 & 101 & - & -   \\
 $c_{66}$ &  88  &  64 & 156 & - & -   
\end{tabular}
\end{table}

\begin{table}
\caption{Phonon frequencies (THz) at the $\Gamma$ and $X$ points of the BZ.}
\label{phontab}
\begin{tabular}{l|cccc}
     &  SC-TB   & DFT & DFT & Expt. \\ 
Mode & this work & Ref.~\onlinecite{Detraux98} &
 Ref.~\onlinecite{Parlinski97} &
Refs.~\onlinecite{Liu87},~\onlinecite{Liu84},~\onlinecite{Cai95} \\ \hline
 & \multicolumn{4}{c}{$\Gamma$ point} \\
$T_{1_u}$ (TO) & { 6.3} & { 8.1} & { 8.5} & { 9.6} \\ 
$T_{2_g}$      & 15.0   & 17.6   &  16.5  & 18.3 \\
$T_{1_u}$ (LO) & -      & 20.1   &  19.7  & 21.1 \\ \hline 
 & \multicolumn{4}{c}{$X$ point} \\
$X_2^-$  & { ~5.1}$i$   &{ ~5.8}$i$  & { ~5.9}$i$ &     \\
$X_5^-$  & { 4.5}       &{ 4.9}      & { 3.5}   & { 5.1} \\
$X_5^+$  & { 5.0}       &{ 8.9}      & 11.7  &     \\
$X_4^-$  & 12.5         & 11.0       & 11.6  &     \\
$X_4^+$  & 18.1         & 17.0       & 16.0  &     \\
$X_2^-$  & 25.0         & 21.0       & 21.0  &     \\ 
\end{tabular}
\end{table}


\end{document}